\documentclass[12pt,a4paper]{article}  
\usepackage{psfrag}   
\usepackage{graphics}  
\usepackage{amssymb,epsfig,amsmath,euscript,array} 
\usepackage{cite}  
 
\newcounter{multieqs}

\newcommand{\be}{\begin{equation}}  
\newcommand{\ee}{\end{equation}}  
\newcommand{\eq}[1]{(\ref{#1})}

\newcommand{\bra}[1]{\langle #1|}  
\newcommand{\ket}[1]{|#1 \rangle}

\newcommand{\bm}[1]{\mbox{\boldmath $#1$}}

\def\bd{\begin{document}}  
\def\ed{\end{document}}  
\def\nn{\nonumber}  
\def\bea{\begin{eqnarray}}  
\def\eea{\end{eqnarray}}  
\let\bm=\bibitem  
\let\la=\label  
  
\def\npb#1#2#3{Nucl. Phys. {\bf{B#1}} #3 (#2)}  
\def\plb#1#2#3{Phys. Lett. {\bf{#1B}} #3 (#2)}  
\def\prl#1#2#3{Phys. Rev. Lett. {\bf{#1}} #3 (#2)}  
\def\prd#1#2#3{Phys. Rev. {D \bf{#1}} #3 (#2)}  
\def\cmp#1#2#3{Comm. Math. Phys. {\bf{#1}} #3 (#2)}  
\def\cqg#1#2#3{Class. Quantum Grav. {\bf{#1}} #3 (#2)}  
\def\nppsa#1#2#3{Nucl. Phys. B (Proc. Suppl.) {\bf{#1A}}#3 (#2)}  
\def\ap#1#2#3{Ann. of Phys. {\bf{#1}} #3 (#2)}  
\def\ijmp#1#2#3{Int. J. Mod. Phys. {\bf{A#1}} #3 (#2)}  
\def\rmp#1#2#3{Rev. Mod. Phys. {\bf{#1}} #3 (#2)}  
\def\mpla#1#2#3{Mod. Phys. Lett. {\bf A#1} #3 (#2)}  
\def\jhep#1#2#3{J. High Energy Phys. {\bf #1} #3 (#2)}  
\def\atmp#1#2#3{Adv. Theor. Math. Phys. {\bf #1} #3 (#2)}  
  
\newcommand{\EQ}[1]{\begin{equation} #1 \end{equation}}  
\newcommand{\AL}[1]{\begin{subequations}\begin{align} #1 \end{align}\end{subequations}}  
\newcommand{\SP}[1]{\begin{equation}\begin{split} #1 \end{split}\end{equation}}  
\newcommand{\ALAT}[2]{\begin{subequations}\begin{alignat}{#1} #2 \end{alignat}\end{subequations}}  
\def\beqa{\begin{eqnarray}}   
\def\eeqa{\end{eqnarray}}   
\def\beq{\begin{equation}}   
\def\eeq{\end{equation}}   
  
\def\N{{\cal N}}  
\def\sst{\scriptscriptstyle}  
\def\thetabar{\bar\theta}  
\def\Tr{{\rm Tr}}  
\def\one{\mbox{1 \kern-.59em {\rm l}}}

\def\a{\alpha}      \def\da{{\dot\alpha}}  
\def\b{\beta}       \def\db{{\dot\beta}}  
\def\c{\gamma}  \def\C{\Gamma}  \def\cdt{\dot\gamma}  
\def\d{\delta}  \def\D{\Delta}  \def\ddt{\dot\delta}  
\def\e{\epsilon}        \def\vare{\varepsilon}  
\def\f{\phi}    \def\F{\Phi}    \def\vvf{\f}  
\def\h{\eta}  
\def\k{\kappa}  
\def\l{\lambda} \def\L{\Lambda}  
\def\m{\mu} \def\n{\nu}  
\def\o{\omega}  
\def\p{\pi} \def\P{\Pi}  
\def\r{\rho}  
\def\s{\sigma}  \def\S{\Sigma}  
\def\t{\tau}  
\def\th{\theta} \def\Th{\Theta} \def\vth{\vartheta}  
\def\X{\Xeta}  
\def\z{\zeta}  
  
\def\cA{{\cal A}} \def\cB{{\cal B}} \def\cC{{\cal C}}  
\def\cD{{\cal D}} \def\cE{{\cal E}} \def\cF{{\cal F}}  
\def\cG{{\cal G}} \def\cH{{\cal H}} \def\cI{{\cal I}}  
\def\cJ{{\cal J}} \def\cK{{\cal K}} \def\cL{{\cal L}}  
\def\cM{{\cal M}} \def\cN{{\cal N}} \def\cO{{\cal O}}  
\def\cP{{\cal P}} \def\cQ{{\cal Q}} \def\cR{{\cal R}}  
\def\cS{{\cal S}} \def\cT{{\cal T}} \def\cU{{\cal U}}  
\def\cV{{\cal V}} \def\cW{{\cal W}} \def\cX{{\cal X}}  
\def\cY{{\cal Y}} \def\cZ{{\cal Z}}

\def\ua{\underline{\alpha}}  
\def\ub{\underline{\phantom{\alpha}}\!\!\!\beta}  
\def\uc{\underline{\phantom{\alpha}}\!\!\!\gamma}  
\def\um{\underline{\mu}}  
\def\ud{\underline\delta}  
\def\ue{\underline\epsilon}  
\def\una{\underline a}\def\unA{\underline A}  
\def\unb{\underline b}\def\unB{\underline B}  
\def\unc{\underline c}\def\unC{\underline C}  
\def\und{\underline d}\def\unD{\underline D}  
\def\une{\underline e}\def\unE{\underline E}  
\def\unf{\underline{\phantom{e}}\!\!\!\! f}\def\unF{\underline F}  
\def\unm{\underline m}\def\unM{\underline M}  
\def\unn{\underline n}\def\unN{\underline N}  
\def\unp{\underline{\phantom{a}}\!\!\! p}\def\unP{\underline P}  
\def\unq{\underline{\phantom{a}}\!\!\! q}  
\def\unQ{\underline{\phantom{A}}\!\!\!\! Q}  
\def\unH{\underline{H}}  
  
\def\As {{A \hspace{-6.4pt} \slash}\;}  
\def\bs {{b \hspace{-6.4pt} \slash}\;}  
\def\Ds {{D \hspace{-6.4pt} \slash}\;}  
\def\ds {{\del \hspace{-6.4pt} \slash}\;}  
\def\ss {{\s \hspace{-6.4pt} \slash}\;}  
\def\ks {{ k \hspace{-6.4pt} \slash}\;}  
\def\ps {{p \hspace{-6.4pt} \slash}\;}  
\def\pas {{{p_1} \hspace{-6.4pt} \slash}\;}  
\def\pbs {{{p_2} \hspace{-6.4pt} \slash}\;}  
  
\def\Fh{\hat{F}}  
\def\Vh{\hat{V}}  
\def\Xh{\hat{X}}  
\def\ah{\hat{a}}  
\def\xh{\hat{x}}  
\def\yh{\hat{y}}  
\def\ph{\hat{p}}  
\def\xih{\hat{\xi}}  
  
\def\psit{\tilde{\psi}}  
\def\Psit{\tilde{\Psi}}  
\def\tht{\tilde{\th}}  
   
\def\At{\tilde{A}}  
\def\Qt{\tilde{Q}}  
\def\Rt{\tilde{R}}  
\def\Nt{\tilde{N}}  
  
\def\at{\tilde{a}}  
\def\st{\tilde{s}}  
\def\ft{\tilde{f}}  
\def\pt{\tilde{p}}  
\def\qt{\tilde{q}}  
\def\vt{\tilde{v}}  
\def\nt{\tilde{n}}  
  
\def\delb{\bar{\partial}}  
\def\bz{\bar{z}}  
\def\bD{\bar{D}}  
\def\bB{\bar{B}}  
  
\def\bk{{\bf k}}  
\def\bl{{\bf l}}  
\def\bp{{\bf p}}  
\def\bq{{\bf q}}  
\def\br{{\bf r}}  
\def\bx{{\bf x}}  
\def\by{{\bf y}}  
\def\bR{{\bf R}}  
\def\bV{{\bf V}}  
  
\def\d{\delta}\def\D{\Delta}\def\ddt{\dot\delta}  
  
\def\pa{\partial} \def\del{\partial}  
\def\xx{\times}  
\def\uno{\mbox{1 \kern-.59em {\rm l}}}    
  
\def\trp{^{\top}}  
\def\inv{^{-1}}  
\def\dag{{^{\dagger}}}  
\def\pr{^{\prime}}  
  
\def\rar{\rightarrow}  
\def\lar{\leftarrow}  
\def\lrar{\leftrightarrow}  
  
\newcommand{\0}{\,\!}      
\def\one{1\!\!1\,\,}  
\def\im{\imath}  
\def\jm{\jmath}  
  
\newcommand{\tr}{\mbox{tr}}  
\newcommand{\slsh}[1]{/ \!\!\!\! #1}  
  
\def\vac{|0\rangle}  
\def\lvac{\langle 0|}  
  
\def\hlf{\frac{1}{2}}  
\def\ove#1{\frac{1}{#1}}  

\def\Box{\square}  
\def\ZZ{\mathbb{Z}}  
\def\CC#1{({\bf #1})}  
\def\bcomment#1{}  
\def\bfhat#1{{\bf \hat{#1}}}  
\def\VEV#1{\left\langle #1\right\rangle}  

\newcommand{\ex}[1]{{\rm e}^{#1}} \def\ii{{\rm i}}

\newcommand{\lrbrk}[1]{\left(#1\right)}
\newcommand{\sfrac}[2]{{\textstyle\frac{#1}{#2}}}

\font\mybb=msbm10 at 12pt
\def\bb#1{\hbox{\mybb#1}}

\font\myBB=msbm10 at 18pt
\def\BB#1{\hbox{\myBB#1}}

\begin{document}  
  
\hfill{{\tt hep-th/0303107}}  
   
\vspace{20pt}  
   
\begin{center}  
  
{\Large \bf BMN operators with vector impurities,\\}   
\vspace{10pt}  
{\Large \bf $\BB{Z}_2$ symmetry and pp-waves }  
\vspace{30pt}  
   
{\bf Chong-Sun Chu$^{a,b}$, 
Valentin V.~Khoze$^{c}$ and 
Gabriele Travaglini$^{c}$}

{\small \em
\begin{itemize}

\item[$^a$]Department of Physics, National Tsing Hua University,
Hsinchu, Taiwan 300, R.O.C.
\item[$^b$] Centre for Particle Theory,
Department of Mathematical Sciences,\\
University of Durham, Durham, DH1 3LE, UK
\item[$^c$] Centre for Particle Theory,
Department of Physics and IPPP,\\
University of Durham, Durham, DH1 3LE, UK
\end{itemize}
}

\vspace{10pt}  
  
Email: {\sffamily \tt chong-sun.chu, valya.khoze,  
gabriele.travaglini@durham.ac.uk }

\vspace{30pt}  
{\bf Abstract}  
  
\end{center}  
We calculate the coefficients of three-point functions 
of BMN operators with two vector impurities. 
We find that these coefficients 
can be obtained from those of the three-point functions of 
scalar BMN operators by  interchanging the coefficient  for
the symmetric-traceless representation with the coefficient for the singlet. 
We conclude that the $\bb{Z}_2$ symmetry 
of the pp-wave string theory is not manifest at the level of
field theory three-point correlators.

\vspace{0.5cm}  
  
\setcounter{page}{0}  
\thispagestyle{empty}  
\newpage


\section{Introduction}

The pp-wave/SYM correspondence of Berenstein, Maldacena and Nastase (BMN)
\cite{BMN}
represents all massive modes of type IIB superstring theory 
in a  plane wave background
in terms of composite BMN operators in  
$\cN=4$ Super Yang-Mills in  four dimensions. 
Until now, most of the calculations on the gauge theory
side of the correspondence were restricted to the BMN operators 
with scalar impurities. 

The goal of the present paper is to extend the study
of correlation functions of scalar BMN operators
\cite{Constable1,CKT,BKPSS,Constable2,CKT2,CK,GK} 
to correlators of vector BMN operators. 
In particular we will address the relevance of a $\bb{Z}_2$ symmetry of the
pp-wave string theory for  the three-point functions of 
vector BMN operators in the gauge theory.
Two-point correlators of BMN operators with vector impurities
have already been considered in \cite{gursoy,beisert,klose}.
We will compute three-point functions of BMN operators
with two vector impurities. 
These three-point functions are essential for the vertex--correlator 
pp-wave duality \cite{Constable1,CKT,CK}.
Our goal is to compare the coefficients
of the three-point functions of vector BMN operators 
with those for scalar BMN operators. 
One would expect that the $\bb{Z}_2$ symmetry of string theory 
in the pp-wave background (explained below) requires the equality  
of these two coefficients. The main result of this paper 
is that this two coefficients are different.%
\footnote{The earlier version of this paper reported an agreement between the
vector and scalar coefficients. This was due to an incorrect handling 
of the compensating terms in the vector BMN operators, last term in 
 \eqref{opndef_symm}.}
Our result is that the vector three-point function \eqref{twoivec}
is related to the scalar three-point function 
\eqref{twoi1} by  exchanging the contribution 
for the symmetric-traceless operator with that of the   singlet. 
This conclusion can also be derived 
from an earlier work of Beisert \cite{beisert}.
From these results it appears that the $\bb{Z}_2$ symmetry 
of the pp-wave string theory is not respected at the level of 
three-point functions of BMN operators with definite scaling dimensions 
in  interacting field theory.

On the string theory side,
the pp-wave background has a bosonic
symmetry of $SO(4)_1 \times SO(4)_2 \times \bb{Z}_2$, where the 
$\bb{Z}_2$ exchanges the action of the two $SO(4)$ groups. 
This symmetry acts quite trivially at the free string level
\cite{z2-1,z2-2}. However, its realisation in the dual 
field theory is not manifest and, therefore, highly non-trivial. 
In the pp-wave/SYM duality, the rotation groups  
$SO(4)_1 \times SO(4)_2$ in the lightcone-gauge string
theory  are mapped to the product of
the  Lorentz (Euclidean) symmetry and the R-symmetry,
$SO(4)_{\rm Lorentz}\times SO(4)_R$, in  the field theory. 
Thus, on the field theory side, the 
$\bb{Z}_2$ factor swaps the action of 
$SO(4)_{\rm Lorentz}$ with $SO(4)_R$.
 A symmetry between 
spacetime and the internal (R-)space
is novel, and might possibly be expected only in the 
large-$N$ double scaling limit. 
The understanding of the $\bb{Z}_2$ symmetry, both in 
interacting string theory \cite{z2-1,z2-2} 
and in field theory, is one of
the most challenging and exciting topics in the pp-wave/SYM duality.

In field theory, the BMN operators that are dual to string excitations
in the first four directions, i.e.~related to the factor $SO(4)_1$,
carry impurities of the form $D_\m Z$ (vector impurities).
Two-point functions and anomalous dimensions of conformal primary 
vector BMN operators have been considered and determined 
in \cite{gursoy,klose}. 
The minimal form of the BMN correspondence is based on the 
mass--dimension type duality relation which
maps the masses of string states to the anomalous dimensions of the
corresponding BMN operators in the gauge theory:
\be 
\label{hh}
{H}_{\rm string} = {H}_{\rm SYM} - J \ .
\ee
This relation has been verified for scalar BMN operators
in the planar limit of SYM perturbation theory
in \cite{BMN,gross,zanon}.
Calculations in the BMN sector of gauge theory at the
nonplanar level were performed in
\cite{KPSS,Constable1,BKPSS,Constable2} also taking into account
mixing effects of planar BMN operators.
The relation was extended
in \cite{ver,gross2,bits2,zhou,Gomis}
to all orders in the effective genus expansion parameter $g_2$.
In \cite{gursoy,klose}
anomalous dimensions of vector BMN operators
were found to be equal to those of scalar BMN operators.
This verifies the
consistency of the $\bb{Z}_2$ symmetry with the
relation \eq{hh}.

However, no further statement has been made so far about the
$\bb{Z}_2$ symmetry beyond the mass-dimension duality
\eq{hh}. 
As we mentioned earlier, we find that  the $\bb{Z}_2$ symmetry 
of the pp-wave string theory is not respected at the level of 
three-point functions of BMN operators with definite scaling dimensions 
in interacting field theory. 

We will first need to carry out a field
theory analysis of the three-point function involving BMN operators
with vector impurities. This part of the analysis is new and 
contains some of the main results of this paper.

Let us recall that in a conformal theory, 
two- and three-point functions of conformal primary
operators are completely determined by conformal invariance. 
One can always choose a basis of {\it scalar} conformal primary operators 
such that the two-point functions take the canonical form:
\begin{equation} 
\label{2pt}
\langle {\cO}_I (x_1) \cO^{\dagger}_J(x_2) \rangle = 
\frac{\d_{IJ}}{(x_{12}^2)^{\Delta_I}} \ ,
\end{equation}
and all the nontrivial information of the
three-point function is contained in the $x$-independent
coefficient $C_{1 2 3}$:
\begin{equation} 
\label{3pt}
\langle \cO_{1}(x_1) \cO_{2}(x_2) \cO^{\dagger}_{3}(x_3) \rangle  =
\frac{C_{123} }
{(x_{12}^2)^{\frac{\D_1+\D_2 -\D_3}{2}}
 (x_{13}^2)^{\frac{\D_1+\D_3 -\D_2}{2}}
 (x_{23}^2)^{\frac{\D_2+\D_3 -\D_1}{2}}} \ ,
\end{equation}
where $x_{ij}^2: = (x_i-x_j)^2$.
Since the form of the $x$-dependence of 
conformal three-point functions is universal,
it is natural to expect that the spacetime independent coefficient $C_{123}$
is related to the interaction of the corresponding three string states
in  the pp-wave background.
Note that, in order to be able to use the coefficients $C_{123}$,
it is essential to work on the SYM side with $\Delta$-BMN operators.
These operators are defined in such a way that they do not mix
with each other (i.e.~have definite scaling
dimensions $\Delta$) and are conformal primary operators.
Conformal invariance
of the $\cN=4$ theory then implies that the two-point correlators
of scalar $\Delta$-BMN
operators are canonically normalized,
and the three-point functions take the simple form \eqref{3pt}.
Defined in this way, the basis of $\Delta$-BMN operators is 
unique and distinct from other BMN bases considered in the literature. 
For two scalar impurities,
this $\Delta$-BMN basis was constructed in \cite{BKPSS}.

However, due to their nontrivial transformation properties under 
the conformal group, 
conformal primary {\it vector} BMN operators 
have in general  more complicated 
two- and three-point functions. 
Thus, a priori, it is not clear whether it is possible (and how) 
to extract in the vector case
a spacetime independent coefficient, similar to the
$C_{123}$ of the scalar correlators,  that can then be compared
with the pp-wave string interaction.  In our opinion, this is one of the
main obstacles in the understanding of how
the pp-wave/SYM duality works for vector impurities and of
the r{\^o}le of the $\bb{Z}_2$ symmetry
beyond the level of the two-point functions in the pp-wave/SYM
correspondence. 
In this paper, we make the observation  
that  in a certain large distance limit, the
two- and three-point correlation
functions for vector BMN operators reduce to the same form as that for 
the scalar case. 
This allows one to
make a direct comparison with  the corresponding
scalar three-point functions.

The paper is organised as follows. In Section 2,
we present the BMN operators with vector impurities and with positive
R-charge. To obtain non-vanishing correlators, one also needs 
to know the conjugate BMN operators, i.e. the BMN
operators with negative R-charge. We construct these operators by employing
a  new conjugation operation which is a product of the usual
hermitian conjugation with the inversion operation. We explain why
this construction is the
most natural one in
the present context. 
An important advantage of our construction is that the
vector BMN operators are orthonormal with respect to the inner
product defined using this conjugation. 
In Section 3 we compute the three-point functions 
involving vector-BMN operators with definite scaling dimensions 
in interacting field theory.

\vspace{0.5cm}
  
\centerline{******}

\bigskip

\centerline{\it Note on notation and conventions}

We write the bosonic part of the 
$\N =4$ Lagrangian as
\be
{\cal{L}} = {2\over g^2} \ \Tr \left( {1 \over 4} F_{\mu \nu}F_{\mu \nu} + 
{1\over 2}(D_\mu \varphi_i) (D_\mu \varphi_i ) 
-{1\over 4} [\varphi_i ,\varphi_j][\varphi_i ,\varphi_j]\right) 
\ , 
\ee 
where $\varphi_i$, $i=1, \ldots , 6$ are the six real scalar fields
transforming under an R-symmetry group $SO(6)$. 
The  covariant derivative is 
 $D_\mu \varphi_i = \partial_{\mu}\varphi_i  - i [A_\mu, \varphi_i ]$, 
where 
$A_\mu = A_\mu^{a} T^a$, and $F_{\mu \nu} = \partial_{\mu}A_{\nu} - 
\partial_{\nu}A_{\mu} - i [ A_{\mu} , A_{\nu}]$.
If we define the complex combinations
\beq
\label{compbas}
\phi^1 = \phi = {\varphi_1 + i \varphi_2 \over \sqrt{2}} \ , \qquad  
\phi^2 = \psi = {\varphi_3 + i \varphi_4 \over \sqrt{2}} \ , \qquad 
\phi^3 = Z  = {\varphi_5 + i \varphi_6 \over \sqrt{2}} \ , 
\eeq
the $\N =4$ Lagrangian can be re-expressed as 
\be
{\cal{L}} = {2\over g^2} \ \Tr \left( {1 \over 4} F_{\mu \nu}F_{\mu \nu} + 
\overline{(D_\mu \phi_I )} (D_\mu \phi^I )\right) 
+ V_F + V_D \ , 
\ee 
where
\beqa
V_F & = & - {2\over g^2}\  \Tr \left( 
[ \phi^I , \phi^J ] [\bar\phi_I ,  \bar\phi_J ] \right) 
\ = \ -2 \ {2\over g^2} \ \Tr \left( Z \phi \bar{Z} \bar{\phi} - 
\phi \bar{\phi}\bar{Z} Z + \cdots \right) \ , 
\\ \nonumber
V_D & = &  {1\over 2} {2\over g^2}\  \Tr \left( [ \phi^I , \bar\phi_I ] 
 [ \phi^J , \bar\phi_J ] \right) \ =\ 
{2\over g^2} \ \Tr \left( Z \bar{Z}Z \bar{Z} - ZZ \bar{Z}\bar{Z}
+ \cdots \right) \ 
\eeqa
are the F-term and D-term of the scalar potential respectively. 
In the last equalities we write only the terms which will be 
relevant for our analysis.
Our $SU(N)$ generators are normalised as
\beq
\Tr \left( T^a T^b \right) = \delta^{ab} \ , 
\eeq
so that, for example,  
\beqa 
\left< Z^{i}_{j}(x) \bar{Z}^{l}_{m}(0) \right> = 
{g^2\over 2} \delta^{i}_{m} \delta^{l}_{j}\Delta (x) 
\ \ , \ \ \Delta (x)= {1\over 4\pi^2 x^2 } \  .
\eeqa
The pp-wave/SYM duality is supposed to hold in the BMN large $N$
double scaling limit,
\be \label{doublel}
J \sim \sqrt{N} \ , \qquad N\to \infty  \ .
\ee
In this limit there remain two free finite dimensionless
parameters \cite{BMN,KPSS,Constable1}:
the effective coupling constant of the BMN sector of gauge theory,
\be \label{lampr}
 \l' = \frac{g^2 N}{J^2} = \frac{1}{(\mu p^+ \a')^2}\
\ee
and the effective genus counting parameter
\begin{equation} \label{gtwo}
g_2 :=\frac{J^2}{N}= 4 \pi g_s (\mu p^+ \a')^2 \ ,
\end{equation}
of Feynman diagrams.
The right hand sides of \eqref{lampr}, \eqref{gtwo} express $\l'$ and
$g_2$ in terms
of the pp-wave string theory parameters.

\bigskip

  
\section{Conformal  primary vector BMN operators} 
Here we will study the BMN operators with 
vector impurities\footnote{CSC and VVK acknowledge an early collaboration with
Michela Petrini, Rodolfo Russo and Alessandro Tanzini on the radial
quantisation method and its applications to vector BMN operators 
discussed in section 2 of this paper.
}.
We will be concerned with  the operators 
\beq
\cO_{\rm vac}^J = {1\over \sqrt{J N_{0}^{J}}}
\Tr Z^J \ , 
\eeq
and, 
for  $\m, \n =1, \ldots, 4$, 
\beq 
\label{opndef}
\cO_{\mu \nu , n}^J =\cC  
\left( \sum_{l=0}^{J}e^{2\pi i nl \over J} \Tr \left[ 
( D_{\mu}Z) Z^l ( D_{\nu}Z)Z^{J-l} \right]+ 
\Tr \left[( D_{\mu}D_{\nu} Z) Z^{J+1}\right]\right) \ + \ \cdots \ ,
\eeq
where we defined 
\beq
\label{defofc}
\cC := {1\over 2 \sqrt{J N_0^{J+2}}}
\ , \qquad  N_0 := {g^2 \over  2}\,{N \over 4\pi^2} \ .
\eeq
The normalisation of the operator $\cO_{\rm vac}^J$ is such that 
its two-point function takes the canonical form \eqref{2pt}
in the planar limit. 
As for the vector BMN operator $\cO_{\mu \nu , n}^J$, it is normalized in
such a way that Eq. \eqref{stringoverl} below holds.
We note that
this choice of normalisation constant $\cC$ 
is different from that\footnote{In particular we have the
same normalisation constant for both cases $n=0$ and $n\neq 0$. 
This 
is related to our prescription for the operator conjugation and 
the definition of the inner product. 
We will explain how this prescription is dictated 
by the pp-wave/SYM correspondence.}
adopted in \cite{klose}. 

The first operator, $\cO_{\rm vac}^J$, 
is a chiral (half-BPS) primary operator, 
and corresponds to the vacuum state of pp-wave string theory. 
For $n \neq 0$,
the second operator, $\cO_{\mu \nu , n}^J$ is a 
non-chiral vector conformal primary BMN operator, and corresponds 
to a string state $\ket{ \a_{n}^{\mu \dagger}\a_{-n}^{\nu \dagger}}$.
Here $\mu$ and $\nu$ are indices of bosonic excitations of the 
first $SO(4)$ in the lightcone pp-wave string theory.%
\footnote{We adopt the convention that 
BMN operators with vector (resp.~scalar) impurities correspond to 
bosonic excitations  of the first (resp.~second)
$SO(4)$ in the lightcone pp-wave string theory.} 
 The operator $\cO_{\mu \nu , n}^J$ has a definite scaling dimension,
$\Delta_{n} = \Delta^{(0)} + \delta_n$, 
which implies that the single-trace expression on the right hand side of
\eqref{opndef} must be accompanied with multi-trace corrections 
(and other mixing effects) at higher orders in $g_2$
\cite{Bianchi,BKPSS}.
The dots on the right hand side  of \eqref{opndef} 
indicate these corrections. These mixing terms are important in general,
but in this paper we will show how to calculate correlation functions
involving operators \eqref{opndef}
without the need of knowing the precise analytical expressions for 
these mixing terms.%

To be more precise, we should distinguish between symmetric-traceless, 
antisymmetric and singlet representations:
\beqa 
\label{opndef_symm}
\cO_{(\mu \nu ) , n}^J &=&\cC  
\left( \sum_{l=0}^{J}e^{2\pi i nl \over J} \Tr \left[ 
( D_{( \mu}Z) Z^l ( D_{\nu )}Z)Z^{J-l} \right]+ 
\Tr \left[( D_{(\mu}D_{\nu )} Z) Z^{J+1}\right]\right) \ + \ \cdots \ 
\\
\label{opndef_antisymm}
\cO_{[ \mu \nu ] , n}^J & = &\cC  
\left( \sum_{l=0}^{J}e^{2\pi i nl \over J} \Tr \left[ 
( D_{[ \mu} Z) Z^l ( D_{\nu ] }Z)Z^{J-l} \right] \right)\ + \ \cdots \ ,
\\
\label{opndef_singl}
\cO_{n}^J & = &\cC  
\left( \sum_{l=0}^{J}e^{2\pi i nl \over J} \Tr \left[ 
( D_{\mu}Z) Z^l ( D_{\mu}Z)Z^{J-l} \right]\right) \ + \ \cdots \ ,
\eeqa
where 
\beqa
\cO_{(\mu \nu )} &= & {1\over 2} \left( 
\cO_{\mu \nu} + 
\cO_{\nu \mu} \right) - {1 \over 4}\  \delta_{\mu \nu}\  \cO_{\rho \rho } 
\ \ , 
\\
\cO_{[\mu \nu ] } &= & {1\over 2} \left( 
\cO_{\mu \nu  } - 
\cO_{\nu \mu  } \right) \ \ . 
\eeqa
Notice that the compensating term 
$\Tr \left[( D_{(\mu}D_{\nu )} Z) Z^{J+1}\right]$ is present only 
in the definition of the symmetric-traceless operator 
in \eqref{opndef_symm} and not in the singlet \eqref{opndef_singl}.
The precise form of the  operators \eqref{opndef_symm}--\eqref{opndef_singl}
is determined by acting with 
supersymmetry transformations on the scalar BMN operators in 
\eqref{opscalar}, and it was 
first obtained in  \cite{beisert} (Eqs.~(B.10), (B.11) and (B.12)),  
which are valid also at finite $J$.
Our operators \eqref{opndef_symm}, \eqref{opndef_antisymm} 
and \eqref{opndef_singl} follow  in  the large-$J$ limit
from those in  \cite{beisert}.
Supersymmetry dictates that the single-trace bosonic operators in 
\eqref{opndef_antisymm} and \eqref{opndef_singl} 
must be accompanied by fermionic bilinears $\sim {g}$ 
and scalar bilinears $\sim g^2$ -- see Appendix B of \cite{beisert}
for the precise form of these terms. 
All these corrections, as well as the multi-trace corrections, 
will not be relevant for the calculation of three-point functions
presented in the following sections, 
hence  we will include them in the dots in 
\eqref{opndef_symm}--\eqref{opndef_singl} and
discard them.

For $n=0$, the operator $\cO_{( \mu \nu ), 0}^J$
is a supergravity translational descendant of the vacuum: 
\bea
\label{sugradesc}
\cO_{( \mu \nu ), 0}^J&=& 
\cC
\left( \sum_{l=0}^{J} \Tr \left[ 
( D_{( \mu}Z) Z^l ( D_{\nu )}Z)Z^{J-l} \right] + 
\Tr \left[( D_{( \mu}D_{\nu )} Z) Z^{J+1}\right] \right)
\nonumber \\ \cr
&=&
{\partial_{( \mu}\partial_{\nu )} \Tr Z^{J+2}\over 2J^{3/ 2}  
\sqrt{N_{0}^{J+2}}}\ .
\eea
This operator is protected, hence its conformal dimension is given by  
the engineering dimension.

We now note that the operators 
$\cO_{\mu \nu , n}^J$ are not orthogonal with respect to the scalar product 
$\langle \cO_{\mu \nu , n}^{J \dagger} (x)\cO_{\rho \sigma , m}^J (y)\rangle$, 
and therefore cannot 
correspond to the (orthonormal) basis 
of string states 
$\ket{ \a_{n}^{\mu \dagger}\a_{-n}^{\nu \dagger}}$
(at least not directly). 
For example,
one has \cite{klose}
for the translational descendant defined in \eqref{sugradesc}, 
\beq
\label{sugranorm1}
\langle \cO_{( \mu \nu ) , 0}^{J \dagger} (x)
\cO_{( \rho \sigma ) , 0}^J (0)\rangle
=
{4 J^2\over (x^2)^{J+4}}{x_{(\mu} x_{\nu )} x_{( \rho} x_{ \sigma )} \over x^4}
\ ,
\eeq
which
is non-zero for $\mu,\nu \neq \rho, \sigma$.
We also note that,  in order to keep the right hand side of 
(\ref{sugranorm1}) finite as $J\to \infty$, an additional factor of  
$J^{-1}$ would be required in the
definition \eqref{sugradesc}
of $\cO_{\mu \nu , 0}^J$  \cite{klose}. 

The right hand side of \eq{sugranorm1} has nothing to do
with an orthonormality of the string states.
We therefore introduce a different  notion of conjugation, 
which will allow a direct  correspondence to string 
(and supergravity) states  defined as  {\it hermitian conjugation} 
followed by an {\it inversion}:%
\footnote{We illustrate the following procedure
for the symmetric-traceless operators
\eqref{opndef_symm}. The extension to the antisymmetric and
singlet  representations is straightforward.}

(i) We define the {\it barred-operator} as
\beqa 
\nonumber
\label{newbarq}
\bar{\cO}_{( \mu \nu ) , n}^J (x) & :=&
\cC \ 
x^{2(\Delta - 2) } \bigg(
\sum_{l=0}^{J}  e^{2\pi i nl \over J}\Tr 
\left[
(J_{( \mu \alpha} \bar{D}_{\a} \ x^2\bar{Z} ) \bar{Z}^l 
( J_{\nu ) \b}\bar{D}_{\b}\ x^2\bar{Z} )\bar{Z}^{J-l} 
\right] 
\\ 
&+&
\Tr \left\{ 
\left[(  J_{( \m \a} \bar{D}_{\a})
(x^2 J_{\n ) \b} \bar{D}_{\b}) x^2 \bar{Z}\right] 
\bar{Z}^{J+1}\right\} \bigg)
\ ,
\eeqa
where $J_{\mu\nu}(x) =\d_{\mu\nu}-2x_{\mu}x_{\nu}/x^2$ 
is the usual inversion tensor, in terms of which the Jacobian 
of the inversion  $x'_{\mu} = x_{\mu} / x^2$ is expressed
$\partial x'_{\mu}  / \partial x_{\nu} = J_{\mu \nu} (x) / x^2 $.
 
{(ii)} We introduce the inner product
\be \label{inner}
\lim_{x\to \infty} \; \langle \bar{\cO_1}(x) \cO_2(0)\rangle
\ee
and, 

{(iii)} propose 
the correspondence between field theory and string theory inner products:
\be \label{inner-corresp}
\lim_{x\to \infty} \; \langle \bar{\cO_1}(x) \cO_2(0) \rangle
\leftrightarrow 
\langle \a_1|\a_2 \rangle,
\ee
where $\ket{\a_\ii}$ is the string state that is in correspondence with the field theory
operator $\cO_\ii$. 

We remark that
the introduction of the barred-operator 
is completely  natural in the context  
of the radial quantisation  of field theory \cite{Fubini}, 
where hermitian conjugation is always accompanied 
by an inversion. Indeed, under inversion  
a scalar field $\cO_{\Delta}(x)$
of conformal dimension $\Delta$ transforms as \cite{mack,osborn}
\beq
\label{scalinv}
\cO_{\Delta} (x) \rightarrow \cO_{\Delta}^{'} (x') = 
x^{2 \Delta}  \cO_{\Delta}(x) \ \ , 
\ \ x_{\m}\to x'_{\m} = {x_{\m} \over x^2}
\ .
\eeq
Differentiating both sides of \eqref{scalinv}
with respect to $x'_{\m}$ we obtain
\beq
\partial_{\m}^{'} \cO_{\Delta}^{'} (x') = 
x^2 J_{\m \n} (x) \partial_{\n}\  [x^{2 \Delta} \cO_{\Delta} (x)]
\ .
\eeq
Combining the action of hermitian conjugation with an inversion, 
we get
\beq
\overline{\partial_{\m}\cO}_{\Delta} (x) = 
x^2 J_{\m \n} (x) \partial_{\n}\  [x^{2 \Delta} \cO_{\Delta}^{\dagger} (x)]
\ , 
\eeq
from which it follows\footnote{
A note on conventions: a bar applied to a composite operator $\cO$
will always mean  hermitian conjugation times an inversion as in
\eqref{newbarq}. For ordinary fields we continue to use 
$\bar{Z} = Z^{\dagger}$.}
 that 
\beqa 
\nonumber
\label{newbar}
\bar{\cO}_{( \mu \nu ), n}^J (x)& =&
\cC
\bigg(
\sum_{l=0}^{J}  e^{2\pi i nl \over J}\Tr 
\left[
(x^2J_{( \mu \alpha} \bar{D}_{\a} \ x^2\bar{Z} ) (x^2\bar{Z})^l 
(x^2J_{\nu )\b}\bar{D}_{\b}\ x^2\bar{Z} )(x^2\bar{Z})^{J-l} 
\right] 
\\ 
&+&
\Tr \left\{ 
\left[( x^2 J_{( \m \a} \bar{D}_{\a})
(x^2 J_{\n ) \b} \bar{D}_{\b}) x^2 \bar{Z}\right] 
(x^2 \bar{Z})^{J+1}\right\} \bigg)
\ ,
\eeqa
which is the free-theory expression for \eqref{newbarq}.

We note that the expression for the string operator
\eqref{opndef} can be more compactly  written as 
\cite{gursoy,klose}
\beq 
\label{opndef2}
\cO_{ \mu \nu  , n}^J =
{\cC \over J}
\sum_{i,j=1}^{J+2 }
e^{2\pi i n(j-i) \over J + 2} 
D_{\m}^{x_i}D_{\n }^{x_j}\
\left.
\Tr \left[  Z (x_1)\cdots Z (x_{J+2}) \right]
\right|_{x_1= \cdots= x_{J+2} = x}
\ .
\eeq
The corresponding expression for the free barred-operator is given then
by 
\beq 
\label{baropndef2}
\bar{\cO}_{\mu \nu , n}^J =
{\cC \over J}
\sum_{i,j=1}^{J+2 }
e^{2\pi i n(i-j) \over J+ 2} 
(x^2 J_{\m \a}D_{\a})^{x_i}
(x^2 J_{\n \b}D_{\b})^{x_j}
\
\left.
\Tr \left[  x_1^2 Z (x_1)\cdots x^2_{J+2} 
Z (x_{J+2}) \right]
\right|_{x_1= \cdots= x_{J+2} = x}
\eeq
We now apply \eqref{newbarq}, or, equivalently
\eqref{newbar}, to 
the protected supergravity operator 
in \eqref{sugradesc}
\beq
\label{sugnew}
\bar{\cO}_{(\mu \nu ), 0}^J= 
{(x^2 J_{( \mu \alpha} \partial_{\a})
(x^2 J_{\nu ) \b}\partial_{\b}) \ \Tr (x^2 \bar{Z})^{J+2}\over 
2 J^{3/ 2} \sqrt{N_{0}^{J+2}}}
\ , 
\eeq
and \eqref{sugranorm1} is now replaced by
the inner product  
\beqa
\label{sugranorm2}
\langle {\bar\cO}_{(\mu \nu ), 0}^J (x)\cO_{(\rho \sigma ) , 0}^J (0)\rangle
 &=& 
(x^2 J_{( \m \a}  \partial_{\a}^{x})(x^2 J_{\n )\b}  \partial_{\b}^{x})
\  \partial_{( \r}^{0}\ \partial_{\s )}^{0} \ (x^2)^{J+2} 
{ \langle \Tr \bar{Z}^{J+2}(x) Z^{J+2}(0) \rangle 
\over
4J^3 N_{0}^{J+2}}
\nonumber \\ \cr
&=& 
\d_{\mu \rho} \d_{\nu \sigma} + \d_{\mu \sigma} \d_{\nu \rho} - 
{1\over 2} \d_{\mu \nu } \d_{\rho \sigma}
\ \ . 
\eeqa
Unlike \eqref{sugranorm1}, 
this expression is  consistent with an operator--supergravity-state 
correspondence. This is the first consistency check of our proposal
\eqref{newbarq} and \eq{inner-corresp}.

We now move on to consider
string states, and compute in the free theory the
two-point function 
$\langle \bar{\cO}_{\mu \nu , n}^J (x){\cO}_{\r \s , m}^J (0)\rangle$
in the limit $x\to \infty$. 
To this end, it is convenient to observe that 
the only terms which survive in this overlap
are the ones where one  derivative operator
originating from the barred operator and one from the unbarred operator 
act on the same propagator, 
$(x^2 J_{\m \a}\partial_{\a}^{x})  \
\partial_{\r}^{y}\ \langle  [x^2\bar{Z} (x)]Z(y) \rangle$.
For these terms 
\beq
\label{imptoimp}
\left.(x^2 J_{\m \a}\partial_{\a}^{x})  \
\partial_{\r}^{y} {x^2 \over (x-y)^2} \right|_{y=0} = 
\left.(x^2 J_{\m \a}\partial_{\a}^{x}) \ x^2 \
{2(x-y)_{\r}\over (x-y)^4} \; \right|_{y=0}
= \;2  \, \delta_{\m \r},
\eeq
where we have used that $\partial_{\a} (x_\r / x^2 )= J_{\a \r} / x^2$, and 
$J_{\m \a} J_{\a \r} = \delta_{\m \r}$.
Keeping this in mind, one easily computes 
in the limit $x\to \infty$,
\beqa
\nonumber
\langle \bar{\cO}_{\mu \nu , n}^J (x){\cO}_{\r \s , m}^J (0)\rangle 
&=& 
\left( {\cC\over J }\right)^{2}  \cdot 
4J^3 \left( {g^2 \over 2 \cdot (4\pi^2)}\right)^{J+2} N^{J+2} \           
(\d_{m,n}\d_{\m \r}\d_{\n \s}  + \d_{m,-n}\d_{\m \s}\d_{\n \r} )
\\ \cr
&=& 
\d_{m,n}\d_{\m \r}\d_{\n \s}  + \d_{m,-n}\d_{\m \s}\d_{\n \r}
\ .
\label{stringoverl}
\eeqa
This result is  again consistent with our operator-string state 
correspondence \eq{inner-corresp}. This is the second, nontrivial  
consistency check of our proposal 
\eqref{newbarq} and \eq{inner-corresp}.
The normalisation chosen in \eqref{opndef} was designed to lead,
on the right hand side of \eqref{stringoverl},
to the product of Kronecker deltas with coefficient equal to 1 .

A few general remarks  are in order:

{\bf 1.} 
In distinction with Eqs.~(29a)--(29d) of \cite{klose}, in our case  
\eqref{stringoverl}, the overlap between  supergravity and string states 
vanishes.

{\bf 2.} On general grounds, conformal invariance requires 
that the two-point function of vector conformal primary operators 
of scaling dimension $\Delta$ should have the form \cite{mack,osborn}:
\beq
\label{conf2}
\langle \cO_{\a \b  , n}^{J \dagger} (x){\cO}_{\r \s , m}^J (0)\rangle = 
{\rm const.} \; {\d_{m,n}J_{\a \r}J_{\b \s}  + \d_{m,-n}J_{\a \s}J_{\b \r} \over
x^{2\Delta}}
\ . 
\eeq
In our approach, we  amputate the coordinate dependence 
on the right hand side of \eqref{conf2}, and contract vector indices with 
(appropriate tensor products of) the inversion tensor $J$, 
thus directly computing
\beq 
\lim_{x \to \infty} \ x^{2 \Delta} 
J_{\m \a} J_{\n \b}  \
\langle {\cO}_{\a \b , n}^{J \dagger} (x){\cO}_{\r \s , m}^J (0)\rangle = 
\d_{m,n}\d_{\m \r}\d_{\n \s}  + \d_{m,-n}\d_{\m \s}\d_{\n \r}
\ , 
\eeq
see our result \eqref{stringoverl}.
We take the limit $x \to \infty$ because $x$ of the barred-operator 
is the inversion of $x'$ and, in the radial quantisation formalism, 
states are obtained from operators at the point $x'=0$.
The corresponding state in radial quantisation would be
\be
\bra{0} (\partial_{\m}^{'} O_{\Delta}^{\dagger '}) (x'=0) \, = \,
\lim_{x \to \infty} 
\bra{0} (x^2 J_{\m \a}(x) 
\partial_{\a})[x^{2\Delta} O_{\Delta}^{\dagger}(x)] \ ,
\ee
which is precisely our definition. 
The two-point functions of vector operators are now correctly normalised, 
and take the canonical form. As a result, they 
are suited for a correspondence with 
the (orthonormal) string theory basis of states.

{\bf 3.}
For the BMN operators with scalar impurities, 
\beq
\label{opscalar}
\cO_{i j , n}^J =
\cC_{\rm scalar}\left(
\sum_{l=0}^{J }e^{2\pi i nl \over J} \Tr \left[
\varphi_{i} Z^l \varphi_{j}Z^{J-l} \right]
\ - \ \d^{ij} \ \Tr (\bar{Z} Z^{J+1})\right)
\ + \ \cdots \ ,
\eeq
one can follow the same  procedure as above, 
and define  the barred-operators as
$\bar{\cO}_{i j , n}^J(x) = x^{2\Delta} \cO^{J \ \dag}_{i j , n} (x)$.
Obviously, whether or not we introduce an inversion for the scalar fields 
is rather irrelevant: 
all the previous results for scalar Green functions are
modified in a straightforward manner 
and the relation \eq{inner-corresp} is verified.
However,  
as we have shown, this leads to  
important differences for vector operators.

{\bf 4.} 
It has been 
argued already in \cite{gursoy,beisert} 
that 
the vector conformal primary BMN operators, i.e. 
$\Delta$-BMN operators with various numbers of vector impurities
are bosonic supersymmetry descendants of the scalar
conformal primary BMN operators. 
This construction has been systematically carried out in 
\cite{beisert}. 
Supersymmetry is important as
it ensures that BMN operators with 
one vector and one scalar impurity \cite{gursoy} or  
two vector impurities \cite{klose} have  exactly the same anomalous dimension
as BMN operators with two scalar impurities, \cite{gursoy,beisert},
in agreement with string theory expectations.



\section{Three-point functions of 
vector conformal primary BMN operators}

Conformal invariance constrains the expression of  three-point functions of 
conformal primary operators. 
For the particular class of three-point functions 
$\langle  \cO_{\r \s, n}^{J_1}(x_1) 
\cO_{\rm vac}^{J_2}(x_2) {\cO}_{\m \n , n}^{J \dagger}(x_3)\rangle$,
involving 
vector conformal primary operators with $J=J_1 + J_2$, one has
\beq
\label{confvector}
\langle 
\cO_{\r \s, n}^{J_1}(x_1) 
\cO_{\rm vac}^{J_2}(x_2) {\cO}_{\m \n , m}^{J \dagger}(x_3) \rangle
=
{F( \r_{n}\s_{-n},\, {\rm vac}|\, \m_m \n_{-m} ; x_{13})
\over
(x_{12})^{\Delta_1 + \Delta_2 - \Delta_3}
(x_{13} )^{\Delta_1 + \Delta_3 - \Delta_2}  
(x_{23})^{\Delta_2 + \Delta_3 - \Delta_1}  
}
\ , 
\eeq
where $x_{ij}=x_i-x_j$, 
$\Delta_i$'s 
are the scaling dimensions of 
$\cO_{\r \s, n}^{J_1}(x_1)$, $ \cO_{\rm vac}^{J_2}(x_2) $ and 
${\cO}_{\m \n , m}^{J }(x_3)$ respectively;
and $F( \r_{n}\s_{-n},\, {\rm vac}|\, \m_m \n_{-m} ; x_{13})$ is a
dimensionless function of $x_{13}$. 
In the quantum theory, 
$\Delta_1 = J_1 + 2 + \delta_n$, 
$\Delta_2 = J_2$, 
$\Delta_3 = J + 2 + \delta_m$, where 
 $\d_m$, $\d_n$  are the anomalous dimensions of 
${\cO}_{\m \n , m}^{J }$, $\cO_{\r \s, n}^{J_1}$.
Therefore
\beqa
 \Delta_1 + \Delta_2 - \Delta_3 &=& \d_n  -\d_m \ , 
\cr 
\Delta_1 + \Delta_3 - \Delta_2 &=& 2(J_1 + 2 ) + \d_n + \d_m \ ,
\cr 
\Delta_2 + \Delta_3 - \Delta_1 &=& 2J_2 + \d_m  -\d_n \ . 
\eeqa
Notice that the anomalous dimensions for 
vector conformal primary operators with one vector and one scalar impurity 
\cite{gursoy}  or with two vector impurities \cite{klose} are the same as for 
the original BMN operators with two scalar impurities \cite{BMN}. 

Conformal invariance requires
$F( \r_{n}\s_{-n},\, {\rm vac}|\, \m_m \n_{-m} ; x_{13})$
to depend on the vector indices $\m$, $\n$, $\r$, $\s$ 
through appropriate tensorial products of the inversion tensor,
 $J(x_{13})\otimes J(x_{13}) $, thus it contains $x$-dependence%
\footnote{See, e.g.~, section III.2 of \cite{Fradkin}.}  
and cannot be compared directly to 
the coefficient $C_{123}$ of the scalar three-point function
\eqref{3pt}, nor with a three-string interaction vertex.
As in the previous section, we propose to consider instead 
the three-point functions involving the barred-operators 
and, moreover, to work in the limit%
\footnote{As before, the limit is a consequence of 
the formalism of radial quantisation. 
We also note that translational invariance, 
broken by radial quantisation, is restored in this limit.}
$x_3\gg x_1,x_2$. 
Using our definition \eqref{newbar} for the barred-operator, 
we will therefore compute 
\beqa
\label{confvector2}
\langle
\cO_{\r \s, n}^{J_1}(x_1) 
\cO_{\rm vac}^{J_2}(x_2) \bar{\cO}_{\m \n , m}^{J }(x_3) \rangle 
&\longrightarrow &
(x_3)^{2\Delta_3}
\langle
\cO_{\r \s, n}^{J_1}(x_1) 
\cO_{\rm vac}^{J_2}(x_2) {\cO}_{\m \n , m}^{J \dagger}(x_3) \rangle 
\cr \cr
&=& 
{C( \r_{n}\s_{-n},\, {\rm vac}|\, \m_m \n_{-m})
\over
(x_{12})^{\d_n - \d_m}
}
\ ,
\eeqa	
for $x_3\to\infty$ (and $x_1$, $x_2$ finite), where
$\cO_{\r \s , n}^{J_1}$ and $\bar{\cO}_{\mu \nu , n}^J$
are given by 
\eqref{opndef} and \eqref{newbarq}.
This is one of the key observation of this paper. 
Now $C( \r_{n}\s_{-n},\, {\rm vac}|\, \m_m \n_{-m})$ 
can be compared directly 
to the scalar three-point function coefficient
$C( k_{n}l_{-n},\, {\rm vac}|\, i_m j_{-m})$, defined below.

The three-point functions of  BMN operators with scalar impurities
\eqref{opscalar} have the form
\beq
\label{sc3ptf}
\langle
\cO_{k l, n}^{J_1}(x_1) 
\cO_{\rm vac}^{J_2}(x_2) {\cO}_{i j , m}^{J \dagger }(x_3) \rangle = 
{C( k_{n}l_{-n},\, {\rm vac}|\, i_m j_{-m})\over
(x_{12})^{\Delta_1 + \Delta_2 - \Delta_3}
(x_{13} )^{\Delta_1 + \Delta_3 - \Delta_2}  
(x_{23})^{\Delta_2 + \Delta_3 - \Delta_1}  
}
\ ,
\eeq
or, introducing the barred-operators and working in the limit
$x_3\to\infty$ (and $x_1$, $x_2$ finite),
\beq
\label{sc3ptfbar}
\langle
\cO_{k l, n}^{J_1}(x_1) 
\cO_{\rm vac}^{J_2}(x_2) \bar{\cO}_{i j , m}^{J }(x_3) \rangle = 
{C( k_{n}l_{-n},\, {\rm vac}|\, i_m j_{-m})\over
(x_{12})^{\d_n - \d_m}}
\ .
\eeq
The expression for   the coefficient  of the 
three-point function for BMN operators with 
two scalar impurities is 
\be 
\label{twoi1}
C( k_{n}l_{-n},\, {\rm vac}|\, i_m j_{-m})
=
C_{123}^{\rm vac}\frac{2\,\sin^2(\pi m y)}{y\, \pi^2 (m^2-n^2/y^2)^2}
\left(\delta_{i(k}\delta_{l)j}\,\,m^2+\delta_{i[k}\delta_{l]j}\,\frac{m n}{y}+
\sfrac{1}{4}\delta_{ij}\delta_{kl}\, \frac{n^2}{y^2}\right) \ , 
\ee
where $y= J_1/J$ is the R-charge ratio, 
$C_{123}^{\rm vac}= \sqrt{JJ_1 J_2} / N$
and the symmetric traceless and  antisymmetric traceless combinations of 
two Kronecker deltas  are defined as
\begin{equation}
\delta_{i(k}\delta_{l)j}= \sfrac{1}{2}(\delta_{ik} 
\delta_{lj} +\delta_{il} \delta_{kj}) 
- \sfrac{1}{4}\delta_{ij}\delta_{kl} \ , \quad
\delta_{i[k}\delta_{l]j}= \sfrac{1}{2}(\delta_{ik} 
\delta_{lj} -\delta_{il} \delta_{kj})  \ .
\label{sdel}
\end{equation}
These results were first obtained 
in the simple case $n=0$ in \cite{CKT}. 
The general expression \eqref{twoi1}
was derived in  \cite{BKPSS}.

We now explain how the computation of the vector three-point functions
proceeds.
In subsection {\bf 3.1} we will describe the free-theory computation, 
and devote  {\bf 3.2} to the planar corrections at one-loop. 
In order to efficiently organise our analysis, we will make a 
step-by-step comparison  with the known computation for  the case of 
scalar impurities. 
More precisely, our strategy will consist in identifying 
the ``building blocks'' which lead to the expression 
\eqref{twoi1} for the coefficient 
$C( k_{n}l_{-n},\, {\rm vac}|\, i_m j_{-m})$
 of the three-point function of scalar BMN operators, 
and comparing them to the corresponding building blocks for the 
case of BMN operators with vector impurities.

\subsection{The calculation in free theory}
Let us briefly review the free theory computation for
the three-point function with 
(complex)
scalar impurities,%
\footnote{For the considerations in free theory presented in  this section, 
we can set all anomalous dimensions equal to zero.} 
say $\phi$ and $\psi$ \cite{Constable1}. 
For calculations with scalars we use the complex basis
\eqref{compbas}, but continue calling the BMN operators
as $ \bar{\cO}_{i j , m}^{J }$ and 
$\cO_{k l, n}^{J_1}$. 

Obviously, to get a nonzero result an impurity in the barred operator 
$ \bar{\cO}_{i j , m}^{J }(x_3)$
must be contracted with an impurity in 
$\cO_{k l, n}^{J_1}(x_1)$ 
and the result boils down to the evaluation of the 
Feynman diagram in Figure 1, which gives
\beq
\label{freescal}
{\rm free-scalar:}  \qquad \left( {g^2 \over 2 } \right)^2 
{1  \over (4\pi^2 x_{31}^2)^2}
P_{\rm free} \ .
\eeq
The  factor $P_{\rm free}$
comes from carefully summing the BMN phase factors 
over all the position of 
$\phi$ and $\psi$
impurities in the operators. 
Its explicit form is given in Appendix B, and will not be needed here. 
When $n=0$, \eq{freescal} is the only contribution to the 
three-point function at the  free level. When $n \neq 0$
the mixing with multi-trace operators must be taken into account
\cite{BKPSS,Bianchi} and will modify even free theory results 
at leading order in $g_2$.
These mixing effects being added to the contributions of Figure 1
lead to the result of
\eq{twoi1} \cite{BKPSS}.

\begin{figure}[ht]
\psfrag{phi}{$\bar{\phi}$}
\psfrag{psi}{$\bar{\psi}$}
\psfrag{k}{$k$}
\psfrag{l}{$l$}
\psfrag{x1}{$x_1$}
\psfrag{x2}{$x_2$}
\begin{center}
{\scalebox{0.7}{
\includegraphics{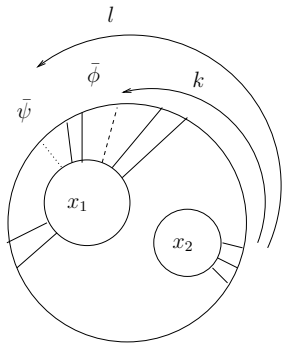}}
}
\end{center}
\caption{Three-point function with scalar impurities. Free diagrams  
contributing to $P_{\rm free}$. 
The labels $k$
and $l$ count the $Z$-lines as indicated (for the diagram drawn above,
$k=2$, $l=4$).  }
\label{fig1}
\end{figure}

\begin{figure}[ht]
\psfrag{phi}{$\overline{D_\m Z}$}
\psfrag{psi}{$\overline{D_\n Z}$}
\psfrag{k}{$k$}
\psfrag{l}{$l$}
\psfrag{x1}{$x_1$}
\psfrag{x2}{$x_2$}
\begin{center}
{\scalebox{0.7}{
\includegraphics{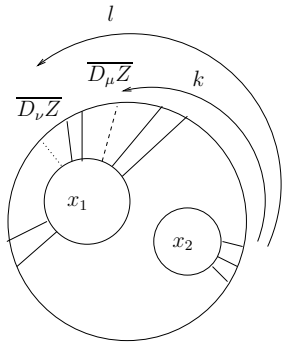}}
}
\end{center}
\caption{Three-point function with vector impurities. Free diagrams  
contributing to 
$P_{\rm free}$. }
\label{fig2}
\end{figure}

We now consider the vector impurity case. 
First, notice that,  in the free theory, 
covariant derivatives can be replaced with simple derivatives. 
The second  key observation is that, 
in the limit we are considering ($x_3 \to \infty$ and $x_1$, $x_2$ finite), 
the only nonvanishing contractions are those where an impurity 
in the operator $ \bar{\cO}_{\m\n , m}^{J }(x_3) $
is  connected to an impurity in 
$\cO_{\rho \sigma, n}^{J_1}(x_1) $.
The result of such contractions 
has been analysed in \eqref{imptoimp}. This observation leads to 
the immediate conclusion that there is only 
one  Feynman diagram contributing  to 
the free vector impurity case (Figure 2). 
The associated phase factor is the same as for the scalar impurity case 
of Figure 1. Therefore the free theory result 
for the vector three-point function is 
given by 
\beq
\label{freevec}
{\rm free-vector:} \qquad   \left( {g^2 \over 2 } \right)^2 
{1   \over (4\pi^2 x_{31}^2)^2}  
P_{\rm free}
\ .
\eeq 
In writing \eqref{freevec} we have taken into account that a factor of 
$2\cdot 2$ from two free contractions of the vector impurities 
(see the right hand side of \eqref{imptoimp}) is precisely cancelled by 
a factor of $(1/2) \cdot (1/2)$ from the normalisation 
of the vector BMN operators.%
\footnote{ 
Notice that the normalisation constant $\cC$ for vector BMN operators is 
half the normalisation of the scalars,
$\cC = ( 1 / 2 )\cdot \cC_{\rm scalar}$.}
Therefore the free result \eqref{freevec} for vector BMN operators 
leads to the same 
result as for the scalars (see \eqref{freescal}).
%

As in the scalar case, there are mixing effects of the 
barred single-trace operator with barred double-trace operators.  
These mixing effects will affect the free-theory contribution of
Figure 2. However, as we argued above, in the region $x_3 \gg x_1,
x_2$, the vector impurities inside a BMN operator are
orthonormal to each other with respect to the inner product
\eq{inner}, and hence behave in the same way as scalar impurities
inside a BMN operator. As a result, 
it is easy to convince oneself that the modifications due to mixing
effects  to the free-theory
three-point function coefficient
are the same for both the scalar and the vector case.
Hence, the free three-point function with vector impurities 
reproduces precisely its counterpart for the case of scalar impurities.

Before concluding this section,  
we would like to discuss further the issue of  mixing. 
The mixing of single-trace BMN operators
with double-trace operators is crucial in order  to obtain conformal 
expressions such as  \eqref{confvector} (or\eqref{confvector2}). 
However, here we are not concerned with deriving  the conformal expression
on the right hand side of  \eqref{confvector}, which must be correct anyway,
as far as the mixing effects are such that we are dealing
with vector conformal primary operators. Our goal is rather to compute
the coefficient of the three-point function with two non-chiral operators, 
$C( \r_{n}\s_{-n},\, {\rm vac}|\, \m_m \n_{-m})$.
At leading order in $g_2$, the only mixing effect which contributes to 
the right hand side of  \eqref{confvector} 
(or \eqref{confvector2})
is the mixing of the barred operator with double-trace operators%
\footnote{To see it immediately, note that the double-trace corrections 
to the single-trace expression for 
a BMN operator is of $\cO (g_2)$, i.e.~suppressed
with $1/N$. This can be compensated by factorising the three-point function 
into a product of two two-point functions. This is possible only 
for the double-trace mixing in the operator $\bar{\cO}$.
}
\cite{BKPSS}.  These mixing effects will affect
not only the free-theory contribution to
$C^{\rm free}( \r_{n}\s_{-n},\, {\rm vac}|\, \m_m \n_{-m})$,  
but also the logarithmic terms $\l' \log x_{13}^2 $ and $\l' \log  x_{23}^2$ 
due to interactions of
the double-trace corrections in $\bar{\cO}_{\m \n , m}^{J }(x_3)$ 
with the BMN operators sitting at $x_1$ and $x_2$. However, it is important to
note that these mixing effects cannot affect
the remaining logarithm, $\l' \log x_{12}^2$  \cite{GK}. 
Hence the coefficient of this logarithm can be computed in planar 
perturbation theory at order $\l'$ without taking into account mixing 
altogether.

Our programme will therefore consist in assuming the conformal form 
(rather than deriving it), and evaluating the terms proportional 
to $\l' \log x_{12}^2$, thus  determining the full coefficient 
of the vector three-point function.
In doing so we are allowed to neglect the double-trace corrections, 
and work directly with the original single-trace BMN expressions.


\subsection{The calculation in the interacting theory  }
The observations made at the end of the last section 
allow us to limit ourselves to the  Feynman diagrams which can generate a 
$\log x_{12}^2$ term. Notice that self-energy corrections 
cannot generate such a $\log x_{12}^2$ dependence, 
and will thus be completely irrelevant for our purposes.

\begin{figure}[ht]
\psfrag{phi}{$\bar{\phi}$}
\psfrag{psi}{$\bar{\psi}$}
\psfrag{k}{$k$}
\psfrag{l}{$l$}
\psfrag{x1}{$x_1$}
\psfrag{x2}{$x_2$}
\psfrag{I}{I}
\psfrag{II}{II}
\begin{center}
{\scalebox{1}{
\includegraphics{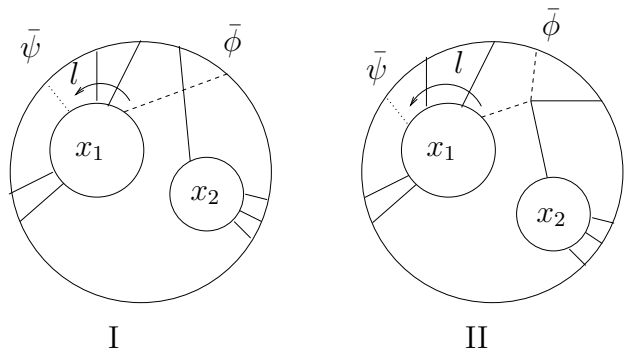}}
}
\end{center}
\caption{Interacting diagrams. Type I: impurity goes across. 
Type II: impurity goes straight. }
\label{fig3}
\end{figure}

\begin{figure}[ht]
\psfrag{phi}{$\bar{\phi}$}
\psfrag{psi}{$\bar{\psi}$}
\psfrag{k}{$k$}
\psfrag{l}{$l$}
\psfrag{x1}{$x_1$}
\psfrag{x2}{$x_2$}
\psfrag{I}{I}
\psfrag{II}{II}
\begin{center}
{\scalebox{1}{
\includegraphics{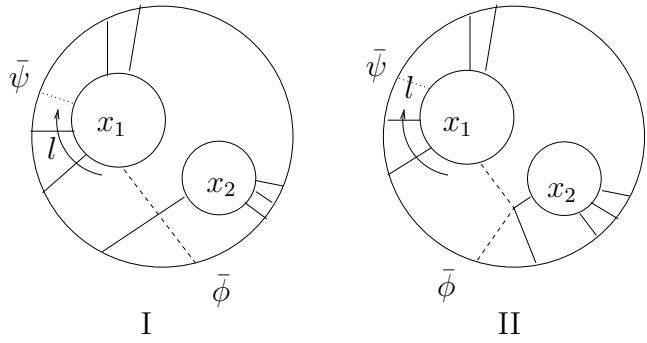}}
}
\end{center}
\caption{Interacting ``mirror'' diagrams. 
}
\label{fig4}
\end{figure}

To begin with, let us recall the situation in the case of scalar impurities.
In that case there are  two diagrams contributing 
to this process, see Figure 3. 
They  come from an
F-term in the Lagrangian, 
$ -V_F = 2 \cdot {2 / g^2} \ \Tr \left( Z \phi \bar{Z} \bar{\phi} - 
\phi \bar{\phi}\bar{Z} Z \right) + \cdots$. 
In the first diagram (type I) the impurity goes across, 
and the diagram comes with coefficient 
$2 \cdot {2 / g^2}$. In the second diagram  
the impurity goes straight (type II), and the diagram  has a  coefficient  
$-2 \cdot {2 / g^2}$. 
 The terms proportional to the $\log x_{12}^2$ resulting  from 
these two diagrams are given by%
\footnote{To keep the formulae as simple as possible, 
we write down only  multiplicative factors of $g^2 / 2$ and
$1 / (4\pi^2 )$ coming from 
the vertices and the propagators involved in the interaction.}
\beqa
{\rm type \ I}-{\rm scalars:}&& +2
\left( {g^2 \over 2 } \right)^3  
P_{I}\  X
\label{typeIF}
\ , 
\\  
{\rm type \ II-scalars:}&& -2 
\left( {g^2 \over 2 } \right)^3 
P_{II}\ X
\label{typeIIF}
\ ,
\eeqa
where the function $X$ is  
\beq
X \, = \, 
- {1
\over 2^8 \pi^6 x_{31}^4}\log x_{12}^2
\, = \, 
{\log |x_{12}|^{-1} \over 8\pi^2} \, \left[ \, \Delta (x_{13} )\, 
\right]^2
\ ,
\eeq
see \eqref{X1234} and \eqref{X} of Appendix A for further details. 
The overall factor $(g^2 / 2)^3$ comes from the insertion of 
one vertex $(2 / g^2 )$, 
and four propagators, $ (g^2 / 2 )^4$. 
The factors of $1 / 4\pi^2$ coming from the propagators are already 
included in the definition of $X$.  Finally, $P_I$ and $P_{II}$ 
are the factors associated with the diagrams of type I and II, 
respectively. Their expressions  are given in Appendix B.

The diagrams drawn in Figure 3 are also accompanied by ``mirror'' diagrams,
where the interaction occurs in the bottom part of the external circle 
(which represents the  barred trace operator) instead 
of  
in the upper part. These diagrams are represented in  Figure 4. 
Their effect is to add to the 
phase factors $P_{I}$ and $P_{II}$ their complex conjugates, 
$\bar{P}_{I}$ and $\bar{P}_{II}$. Finally, there are also the diagrams 
where the interaction involves the impurity  $\psi$ instead of $\phi$. 
The net effect of these diagrams is to double up each phase factor, so that 
and amounts to replacing $P_I$ and $P_{II}$ in \eq{typeIF} and 
\eq{typeIIF} respectively by $2( P_{I}+ \bar{P}_{I})$ and 
$ 2( P_{II}+ \bar{P}_{II})$.
Notice that 
\eqref{typeIF} and \eqref{typeIIF}
must be compared to the free result, 
which was computed in \eqref{freescal}.

We have now assembled the building blocks Eqs.~\eqref{typeIF}, \eqref{typeIIF}
for deriving the formula \eqref{twoi1} for the case of 
different scalar impurities ($i\neq j$, $k\neq l$).
In fact, the derivation of \eqref{twoi1} follows immediately 
by tracking the $\log x_{12}^2$ terms. 
For the case of same impurities, one cannot use the complex basis 
in order to derive \eqref{twoi1}, but a straightforward modification 
of the above  shows that \eqref{twoi1} holds.

Having identified the building blocks for the
scalar-impurities calculation, we are finally ready to study 
the vector-impurities interacting case.

\begin{figure}[ht]
\psfrag{phi}{$\overline{D_\m Z}$}
\psfrag{psi}{$\overline{D_\n Z}$}
\psfrag{k}{$k$}
\psfrag{l}{$l$}
\psfrag{x1}{$x_1$}
\psfrag{x2}{$x_2$}
\psfrag{(1)}{(1)}
\psfrag{(2)}{(2)}
\psfrag{(3)}{(3)}
\begin{center}
{\scalebox{0.6}{
\includegraphics{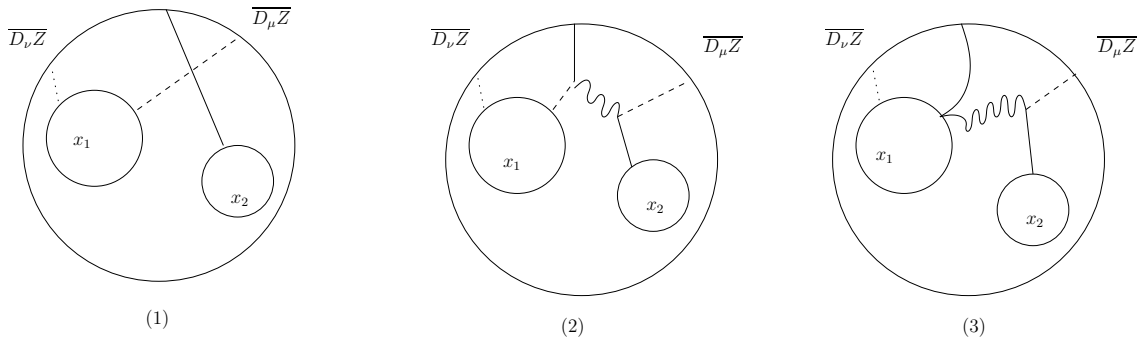}}
}
\end{center}
\caption{Interacting vector diagrams: type I. 
}
\label{fig5}
\end{figure}

\begin{figure}[ht]
\psfrag{Z}{$Z$}
\psfrag{Zc}{$\overline{D_\m Z}$}
\psfrag{Za}{$\overline{D_\n Z}$}
\psfrag{Zb}{$\overline{Z}$}
\psfrag{k}{$k$}
\psfrag{l}{$l$}
\psfrag{x1}{$x_1$}
\psfrag{x2}{$x_2$}
\psfrag{(1)}{(1)}
\psfrag{(2)}{(2)}
\psfrag{(3)}{(3)}
\psfrag{(4)}{(4)}
\psfrag{(5)}{(5)}
\psfrag{(6)}{(6)}
\begin{center}
{\scalebox{0.6}{
\includegraphics{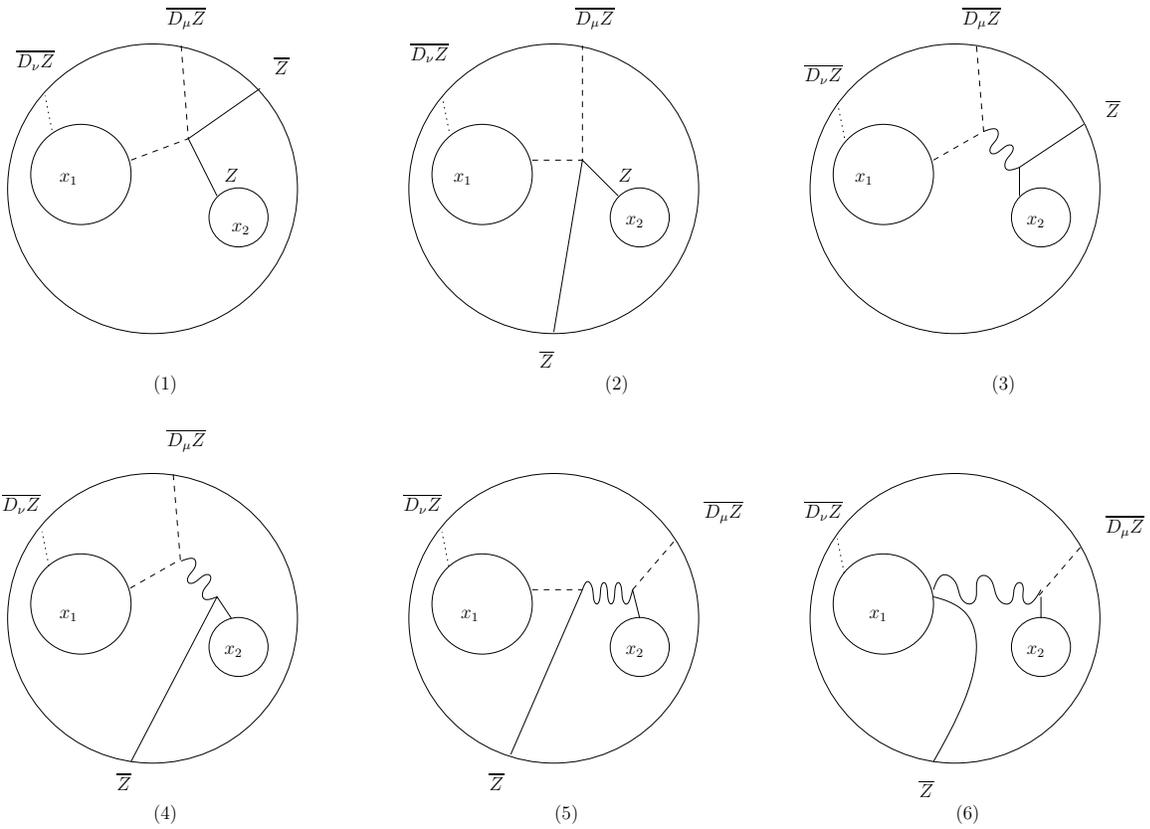}}
}
\end{center}
\caption{Interacting  vector diagrams: type II. 
}
\label{fig6}
\end{figure}

The diagrams where the interaction does not include either of the impurities  
cancel among each other in
both the scalar and  the vector cases.
We are thus left with only two classes of diagrams, 
in complete analogy with the scalar case: 
in the diagrams of the first class, represented in Figure 5, 
the impurity goes across (type I), 
whereas for those in the second class, in Figure 6, 
the impurity goes straight (type II). From 
these diagrams it follows immediately that 
the phase factor associated 
with type I (II) vector diagrams is the same 
as for the corresponding diagrams of type I (II) for scalars.
To establish the $\bb{Z}_2$ symmetry we only need 
to compare the coefficients of the scalar and vector diagrams. 

Let us have a closer look at the diagrams of type I
(impurity goes across).
The first diagram in Figure 5
comes  from a  D-term in the Lagrangian, 
$- V_D= {2 / g^2} \ 
\Tr \left( Z Z\bar{Z} \bar{Z}\right)+ \cdots$. 
Importantly, it has the same sign of the  
F-term contributing to the same class of diagram for the scalar case, 
$-V_F = 2 \cdot {2 / g^2} \ \Tr \left( Z \phi \bar{Z} \bar{\phi} \right)
+ \cdots$.
Its contribution is 
\beq
\label{52}
+  \left( {2 \over g^2}\right)
\left( {g^2 \over 2 } \right)^4 P_{I}
\ X 
\ . 
\eeq
The second diagram in Figure 5 is evaluated in \eqref{Hvanish}
Appendix A and gives a vanishing contribution. 
The third diagram  is the gluon emission from the impurity, 
and comes from the commutator term in the covariant derivative impurity
in  $\cO_{\r \s , n}^{J_1} (x_1 )$.
This diagram is also evaluated in \eqref{Ydiag} of  
Appendix A, 
and the result is:  
\beq
\label{gluon_emissi_I}
+3 
\left( {2 \over g^2}\right)^2\left( {g^2 \over 2 } \right)^5  
P_{I} \ X
\ . 
\eeq
The total answer for the diagrams of type I 
is therefore equal to
\beq
\label{typeIV}
{\rm type \ I-vector:} \qquad
+4\cdot 2 \cdot {1\over 4}  \left( {g^2 \over 2 } \right)^3 
P_{I} \  X \ . 
\eeq
In writing \eqref{typeIV} we have multiplied the sum of 
\eqref{52} and \eqref{gluon_emissi_I} by a factor of $2$ 
from the free contraction of the impurity which does not interact, 
and a factor of $1/4$ from the normalisations of the vector BMN operators.
As in the scalar case, the inclusion of mirror diagrams 
and of the diagrams where the interaction occurs where 
the other impurity is located, 
amounts to replacing  $P_{I} $ in \eqref{typeIV} with  
$2(P_{I} + \bar{P}_{I})$. 
%
In conclusion,  the coefficient 
of the $\log x_{12}^2$ term 
\eqref{typeIV}   arising 
from the  diagrams of type I 
precisely coincides with the corresponding  
coefficient for type I diagrams for 
the scalar three-point functions, 
\eqref{typeIF}.


We now consider the diagrams of type II (Figure 6).
The first and second one originate from the two terms contained in   
$- V_D= {2 / g^2} \ 
\Tr \left(  ZZ \bar{Z}\bar{Z} - Z \bar{Z}Z \bar{Z}
\right)+ \cdots$ respectively, 
and have opposite signs.
The second diagram carries a symmetry factor 2 compared to the first and
their spatial dependence is the same. 
Their combined result is equal to 
\beq
\label{55}
(1-2) \left( {2 \over g^2}\right) \left( {g^2 \over 2 } \right)^4 
P_{II}\ X 
\ . 
\eeq
The third and fourth diagram come with opposite signs 
and have the same spatial dependence, 
therefore their net contribution vanishes.
The fifth diagram vanishes by itself, as the second diagram in 
Figure 5.   
The sixth diagram  follows from a  contribution 
from the commutator term 
in the covariant-derivative impurity present in 
$\cO (x_1)$. The sign of this diagram is opposite to 
to the similar one of type I
(the third in Figure 5),  however this time it comes with phase factor 
$P_{II}$. Its contribution is  therefore equal to 
\beq
\label{gluon_emissi_II}
-3  \left( {2 \over g^2}\right)^2
\left( {g^2 \over 2 } \right)^5
 P_{II}\ X 
\ . 
\eeq
The final  result for the diagrams of type II is
\beq
\label{typeIIV}
{\rm type \ II-vector:} \qquad
-4 \cdot 2 \cdot {1\over 4} \left( {g^2 \over 2 } \right)^3 
P_{II} \  X \ . 
\eeq
As before, in writing the result \eqref{typeIIV} 
we have multiplied the sum of 
\eqref{55} and \eqref{gluon_emissi_II} by a factor of $2$ 
from the free contraction of the non-interacting 
impurity, and a factor of $1/4$ from the normalisations 
of the two vector BMN operators.
Including  mirror diagrams 
and  the diagrams where the interaction occurs where 
the other impurities is located 
amounts to replacing  $P_{II} $ in \eqref{typeIIV} with  
$2(P_{II} + \bar{P}_{II})$.
Therefore, we  conclude that 
the coefficient of the $\log x_{12}^2$ term \eqref{typeIIV} 
from the  diagrams of type II 
precisely matches the corresponding coefficient from 
type II diagrams for the scalar three-point functions, 
\eqref{typeIIF}.


For the vector BMN operators in the antisymmetric and in the singlet 
representations, where the compensating term 
$\Tr \left[( D_{(\mu}D_{\nu )} Z) Z^{J+1}\right]$ is not present, 
the diagrams of type I and II give the full answer. 
Defining 
\beq
P_{m,n}:= 2( P_{I}  + \bar{P}_{I}- P_{II}  - \bar{P}_{II}) = 
  - {8 m \over {m - n/  y}} \sin^2 \pi m y
\ , 
\eeq
the coefficients of the three-point function for the antisymmetric and 
singlet representations are respectively expressed in terms of the 
combinations
\beqa
P_{m,n} - P_{-m,n} &=&  - 8\sin^2 \pi m y \ 
{ 2\ mn /  y \over {m^2 - n^2 /  y^2}} 
\ , 
\\
P_{m,n} + P_{-m,n} &=&  - 8\sin^2 \pi m y \ 
{ {2m^2 } \over {m^2 - n^2 /  y^2}} 
\ .
\eeqa
For vector BMN operators in the symmetric-traceless representation, however, 
the contribution coming from the compensating term in \eqref{opndef_symm}
are important and must be included. In Figure 7 we draw the corresponding 
Feynman diagrams which are associated with a  phase factor equal to 
1.

\begin{figure}[ht]
\psfrag{phi}{$\overline{D_\m Z}$}
\psfrag{psi}{$\overline{D_\n Z}$}
\psfrag{Dr Ds}{$D_{\rho}D_{\sigma} Z$}
\psfrag{k}{$k$}
\psfrag{l}{$l$}
\psfrag{x1}{$x_1$}
\psfrag{x2}{$x_2$}
\psfrag{Z}{$Z$}
\psfrag{(1)}{(1)}
\psfrag{(2)}{(2)}
\psfrag{(3)}{(3)}
\begin{center}
{\scalebox{0.6}{
\includegraphics{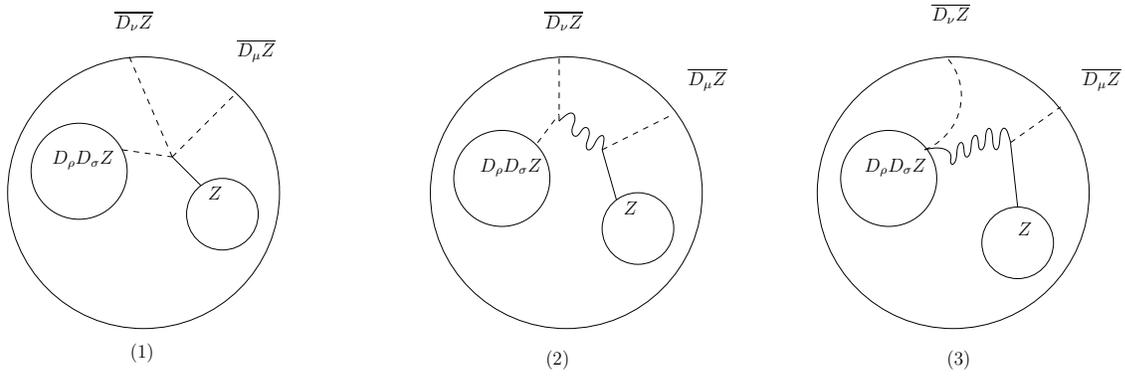}}
}
\end{center}
\caption{Diagrams from the compensating term, with phase factor 1. 
}
\label{fig7}
\end{figure}

The sum of the first and the second diagram (gluon interaction)
in Figure 7 gives a contribution 
\beq
\label{61}
2\cdot {1\over 4}  
\left( {2 \over g^2}\right)^2\left( {g^2 \over 2 } \right)^5  
X
\ . 
\eeq
The third diagram in Figure 7 is the gluon emission from the impurity, 
and gives a contribution
\beq
\label{comp_gluon_emiss}
3\cdot 2 \cdot {1\over 4} 
\left( {2 \over g^2}\right)^2\left( {g^2 \over 2 } \right)^5  
X
\ . 
\eeq
In the expressions \eqref{61} and \eqref{comp_gluon_emiss}
we have included a factor of $1/4$ from the normalisations of 
the vector BMN operators. 
In addition to  the diagrams in Figure 7, 
there are diagrams which are obtained from them  by pulling 
down the upper right leg. 
They  come with a relative factor $- \bar{q}^{J_1 + 1}$, where
$q = \exp 2 \pi i m / J$, and the minus sign comes from pulling 
down the leg.  
We also have to include mirror diagrams, 
and the diagrams with  $\m$ and $\n$ interchanged
(which double up the result).

When $n=0$, the operator at $x_1$ does not in fact contain gluons, 
as is clear from \eqref{sugradesc}, hence the total gluon emission diagrams 
have to cancel for  $n=0$.
This does happen since the gluon emission in the diagrams 
containing the compensating term 
precisely cancels%
\footnote{Notice that the total phase factor 
for the diagrams containing the compensating term is, for large $J$,  
$2\cdot \left( 2 - q^{J_1 } - \bar{q}^{J_1 } \right) 
\, = \, 8 \sin^2 \pi m y
\, = \,  -P_{m,0} $.} 
the sum of the contributions 
in  \eqref{gluon_emissi_I} and  \eqref{gluon_emissi_II} 
at $n=0$.

The complete result for the symmetric-traceless representation is obtained 
by adding together the contributions arising from 
the diagrams of type I and II and all the contributions from 
the compensating term. 
The coefficients of the three-point function for the symmetric-traceless
representation is therefore expressed in terms of
\beq
P_{m,n} + P_{-m,n} - (P_{m,0} + P_{-m,0}) = 
-8\sin^2 \pi m y \ 
{ 2 \ n^2 / y^2  \over {m^2 - n^2 / y^2}} 
\ .
\eeq

Summarising, 
if we first ignore  the compensating terms, 
the building blocks 
\eqref{typeIV} and  \eqref{typeIIV}
for deriving the expression for the coefficient
$C( \r_{n}\s_{-n},\, {\rm vac}|\, \m_m \n_{-m})$
of the vector three-point function 
are  precisely the same
as the building blocks 
\eqref{typeIF}, \eqref{typeIIF} for 
the scalar three-point function coefficient
$C( k_{n}l_{-n},\, {\rm vac}|\, i_m j_{-m})$ 
(which in turn lead to the expression 
\eqref{twoi1}). 
However, in the vector case a compensating term is present 
in the symmetric-traceless representation, 
whereas in the scalar case the compensating term 
affects instead the singlet operator. 
Therefore, it follows that 
$C( \r_{n}\s_{-n},\, {\rm vac}|\, \m_m \n_{-m})$
is given by 
\be 
\label{twoivec}
C( \r_{n}\s_{-n},\, {\rm vac}|\, \m_m \n_{-m})
=
C_{123}^{\rm vac}\frac{2\,\sin^2(\pi m y)}{y\, \pi^2 (m^2-n^2/y^2)^2}
\left(\delta_{\m(\r}\delta_{\s)\n}\,\, 
\frac{n^2}{y^2} + 
\delta_{\m[\r}\delta_{\s]\n}\,\frac{m n}{y}+
\frac{1}{4}\delta_{\m\n}\delta_{\r\s}m^2\, \right) \ .
\ee
This is one of the principal results of this paper. We
note that \eqref{twoivec} agrees with the expression proposed
earlier in \cite{beisert}.%
\footnote{In the first version of this paper, 
contributions coming from the compensating term 
for the symmetric-traceless representation 
(the last term in the right-hand side of \eqref{opndef_symm})
were overlooked, 
which resulted in an incorrect 
expression.} 
The vector three-point function \eqref{twoivec}
is related to the scalar three-point function 
\eqref{twoi1} by simply interchanging the contribution 
for the symmetric-traceless 
with that of the   singlet. 
Therefore,  the $\bb{Z}_2$ symmetry 
of the pp-wave string theory is not respected at the level of 
three-point functions of the BMN operators with definite scaling dimension 
in interacting field theory.


\vspace{0.5cm}
\centerline{******}

In this paper, we have introduced a new notion of 
conjugation to define 
BMN operators with negative R-charge. 
The new conjugation is a composition of the usual hermitian conjugation 
with an inversion, and is entirely
consistent with the spirit of radial quantization. Using this
conjugation, we introduced a new inner product for the BMN operators, 
which is relevant for the pp-wave/SYM correspondence 
and maintains the orthonormality of the string states.
We computed three-point functions for BMN operators with two vector impurities.
The corresponding coefficient is given in \eqref{twoivec}. 
This expression should be contrasted with the result 
in the  scalar case, \eqref{twoi1}. 
Contrary to expectations based on the $\bb{Z}_2$ invariance
in pp-wave string theory, these two expressions are different.


\section*{Acknowledgements} 
Particular thanks go to    
Niklas Beisert and George Georgiou for discussions
and comments on this paper. 
We would also like to thank Gleb Arutyunov, Massimo Bianchi, 
 Michela Petrini, Giancarlo Rossi, Rodolfo Russo, 
Yassen Stanev and Alessandro Tanzini
for useful   conversations. 
CSC would like to thank Pinpin Chen for interesting  discussions.
We acknowledge grants from the Nuffield foundation  
and PPARC of UK, and NSC and NCTS of Taiwan.
  
\section*{Appendix A:
Evaluation of the diagrams}

In the computation of the scalar three-point functions we define 
the function   
\beq
\label{X1234}
 X_{1234} = \int d^4z \ 
\Delta (x_1 - z) \Delta (x_2 - z) \Delta (x_3 - z) \Delta (x_4 - z) 
\ \ .
\eeq
$X_{1234}$ develops  a $\log x_{12}^2$ term  $X$ 
as $x_1$ approaches $x_2$,
which is given by  
\beq
\label{X}
X:= \left. X_{1234}\right|_{x_3 = x_4} = 
- {1 \over 2^8 \pi^6} 
 \left(  {1\over x_{14}^2 x_{24}^2} \right)\log x_{12}^2
\ .
\eeq
One can also derive directly from \eqref{X1234} useful expressions for
the derivatives of $X$, 
\beq
\left( 
{\partial \over \partial x_4^\a } X_{1234} \right)_{x_3 = x_4} = 
{1\over 2}  \left( - {1 \over 2^8 \pi^6}\right)  
\log x_{12}^2 \ 
{\partial \over \partial x_4^\a}
\left(  {1\over x_{14}^2 x_{24}^2} \right)
\ , 
\eeq
\beq
\label{jj}
\left( 
{\partial^2 \over \partial x_4^\a  \partial x_4^\b} 
X_{1234} \right)_{x_3 = x_4} = 
{1\over 6}  \left( - {1 \over 2^8 \pi^6}\right)
\log x_{12}^2 \ 
{\partial^2 \over \partial x_4^\a  \partial x_4^\b} 
\left(  {1\over x_{14}^2 x_{24}^2} \right)
\ ,
\eeq
where, on the right-hand side of \eqref{jj} 
derivatives are taken as if $\a \neq \beta$, i.e.~$\partial_{\a}x^{\b}=0$
rather than $\d_{\a}^{\b}$.

In the evaluation of all the diagrams with an insertion of 
quadrilinear term in the scalars coming from $-V_D$
we also made use of the following relations: 
\beqa
\left[ J_{\m \a}(x_3)\partial_{\a}^{x_3} x_3^2 \ 
\partial_{\r}^{x^1} \ X_{1234}\right]_{x_3 = x_4}  
&\longrightarrow& \ \ \delta_{\m \r} X \ , 
\\ \cr
\left[ J_{\m \a}(x_4)\partial_{\a}^{x_4} x_4^2 \ 
 \partial_{\r}^{x^1} \ X_{1234}\right]_{x_3 = x_4}  
&\longrightarrow &\ \ \delta_{\m \r} X \ , 
\eeqa
where equality with the right hand sides holds 
for the $\log x_{12}^2$ terms,  in the limit
$x_{12} \to 0$ and $ x_3 \to \infty$, and we used
\beq
\partial_{\a} {x_{\b} \over x^2} = {J_{\a \b}(x) \over x^2}
\ .
\eeq
The covariant derivative interaction term, which participates in 
the third diagram  of Figure 5 and in the sixth diagram of Figure 6, 
is proportional to 
$(\partial^{x_2}_{\mu}  - \partial^{x_3}_{\mu})Y_{123}$, where
\beq
\label{Ydef}
Y_{123} = \int d^4z \ 
\Delta (x_1 - z) \Delta (x_2 - z) \Delta (x_3 - z) 
\ .
\eeq
It is easy to realise that, as $x_{12} \to 0$, the function $Y_{123}$
contains a logarithmic term given by
\beq
\label{pp}
\left. Y_{123}\right|_{x_{12} \to 0}  =  
 - {1 \over 2^6 \pi^4}  
 \log x_{12}^2 \ 
{1 \over x_{13}^2} 
\ .
\eeq
One also needs the following expression for the $ \log x_{12}^2$ term in the
 first derivative of $Y$:
\beq
\label{ii}
\left( {\partial \over \partial x_1^\a } Y_{123}\right)_{x_{12} \to 0}  = 
- {1 \over 2^7 \pi^4}  
 \log x_{12}^2 \ 
{\partial \over \partial x_1^\a } {1 \over x_{13}^2} 
\ .
\eeq
We note  that \eqref{ii} is obtained from \eqref{Ydef} rather than 
by differentiating  \eqref{pp}. 

To compute the diagram, we also used that, as $x_{12} \to 0$,  
\beq
\label{Ydiag}
[J_{\m \b}(x_3 ) \partial_{\b}^{x_3}\   x_{3}^2 ]\ 
( \partial_{\r}^{x_2} - \partial_{\r}^{x_3}) Y_{123} 
\longrightarrow
3 X \ \d_{\m \r} 
\ .
\eeq
The contribution of the second diagram in Figure 5 (gluon interaction)
is encoded in the function $H$ defined by
\beq
H_{14,23} = 
(\partial_{\m}^{x_1} -  \partial_{\m}^{x_4})
(\partial_{\m}^{x_2} -  \partial_{\m}^{x_3})
\int d^4 z \  d^4 t\ \ \D (x_1 - z ) \D (x_4 - z )   
  \D (x_2 - t ) \D (x_3 - t )                 
\D (z - t )
\ , 
\eeq
which can be evaluated with the useful relation  proved in 
\cite{BKPSS}
\beq
\label{H}
{H_{14,23} \over \D_{14}\D_{23} }= 
X_{1234} \left( {1\over \D_{12}\D_{43} }-{1\over \D_{13}\D_{24} }\right) + 
G_{1,23} - G_{4,23}+G_{2,14}-G_{3,14}
\ \ ,
\eeq
where $\D_{ij} = \D (x_i - x_j)$ and 
\beq
G_{i,jk}= Y_{ijk}\left( {1 \over \D_{ik}} - {1 \over \D_{ij}}\right)
\ .
\eeq
We can recast \eqref{H} as
\beqa
H_{14,23} &=& -X_{1234} {\Delta_{14}\Delta_{23}\over \Delta_{13}\Delta_{24}} 
\ +  \
\left( {Y_{123} \over \D_{13}} +  {Y_{124} \over \D_{24}} \right)
\D_{14}\D_{23} \ + \ \cdots 
\nonumber \\ \cr
&= & H_{I} \ + \ H_{II} + \ \cdots 
\ \ ,
\eeqa
where the dots stand for terms which either vanish or do not contain the  
$\log x_{12}^2$.
The vanishing of the second diagram in Figure 5 stems from the fact that
\beq
\label{Hvanish}
[ J_{\m \a}(x_3) \partial_{\a}^{x_3} x_3^2] \ \partial_{\m}^{x_1} H_{I}  
 = 
- [J_{\m \a}(x_3) \partial_{\a}^{x_3} x_3^2] \ \partial_{\m}^{x_1} H_{II} 
\ ,
\eeq
where the equality holds for the $\log x_{12}^2$  terms we are looking at, 
and in the limit $x_3 = x_4 \to \infty$.

\section*{Appendix B:
Summing the BMN phase factors}
We report here the expressions for the 
coefficients $P_{\rm free}$, $P_{I}$ and $P_{II}$ 
which arise after summing over the BMN phase factors 
in the free-theory diagrams and in the diagrams of type I and II, 
respectively. Defining 
\beq
q = e^{2\pi i m  / J} \ , \qquad 
q_1 = e^{2\pi i n /  J_1} \ ,
\eeq
we have, for the free case, 
\beq
\label{pfree}
 P_{\rm free} = \sum_{k,l=0}^{J_1} \ (\bar{q} q_1)^{l-k} + 
\sum_{l=0}^{J_1} \   (\bar{q} q_1)^{0} \ = \ 
{J^2 \over \pi^2} {\sin^2 \pi m y \over (m-  n /  y)^2}
\ + \ \cO(J) \ ,
\eeq
where the last equality holds in the BMN limit, and $y=J_1 / J$.
The expressions for $P_{I}$ and $P_{II}$ are given by 
\beq
P_{I} = \sum_{l=0}^{J_1}\  (\bar{q} q_1)^{l}
\ \bar{q} \ , \qquad 
P_{II} = \sum_{l=0}^{J_1} \ (\bar{q} q_1)^{l}
\ . 
\eeq
The effective coefficient which  multiplies
the $\log x_{12}^2$ term in  the three-point function, 
both in the scalar and in the vector case,  is
\beqa
\label{totalpf}
2(P_{I} + \bar{P}_{I}) - 2(P_{II} + \bar{P}_{II}) 
&=& 
2\ \sum_{l=0}^{J_1} \ (\bar{q} q_1)^{l}\  (\bar{q} - 1 )
\ + \  {\rm c.c.} 
\nonumber \\ \ \cr
&=&
 - {8 m \over {m - n/  y}} \sin^2 \pi m y
\ . 
\eeqa
Again, the last equality holds in the BMN limit. Notice that 
\eqref{totalpf} is of $\cO( 1 / J^2 )$ compared with $P_{\rm free}$ in 
\eqref{pfree}, as it should. This, together with a factor $g^2 N$ of 
with the planar one-loop contribution to the three-point function,  
reconstruct the effective Yang-Mills coupling constant $\l' = g^2 N / J^2$, 
which is kept  fixed in the BMN limit.


\begin{thebibliography}{99}  
  


\bibitem{BMN}  
D.~Berenstein, J.~M.~Maldacena and H.~Nastase,  
{\it ``Strings in flat space and pp waves from N = 4 super Yang Mills,'' } 
JHEP {\bf 0204} (2002) 013,  
{\tt hep-th/0202021}.  
  
\bibitem{Constable1}  
N.~R.~Constable, D.~Z.~Freedman, M.~Headrick, S.~Minwalla, L.~Motl,  
A.~Postnikov and W.~Skiba,   
{\it ``PP-wave string interactions from perturbative Yang-Mills theory,''  }
JHEP {\bf 0207} (2002) 017, {\tt hep-th/0205089}.  



\bibitem{CKT}  
C.~S.~Chu, V.~V.~Khoze and G.~Travaglini,  
{\it ``Three-point functions in N = 4 Yang-Mills theory and pp-waves,''  }
JHEP {\bf 0206} (2002) 011,  {\tt hep-th/0206005}.  

\bibitem{BKPSS}
N.~Beisert, C.~Kristjansen, J.~Plefka, G.~W.~Semenoff and M.~Staudacher,
{\it ``BMN correlators and operator mixing in N = 4 super Yang-Mills theory,''}
Nucl.\ Phys.\ B {\bf 650} (2003) 125,
{\tt hep-th/0208178}.


\bibitem{Constable2}
N.~R.~Constable, D.~Z.~Freedman, M.~Headrick and S.~Minwalla,
{\it ``Operator mixing and the BMN correspondence,''}
JHEP {\bf 0210} (2002) 068,
{\tt hep-th/0209002}.

\bibitem{CKT2}
C.~S.~Chu, V.~V.~Khoze and G.~Travaglini,
{\it ``pp-wave string interactions from n-point correlators of BMN operators,''}
JHEP {\bf 0209} (2002) 054,
{\tt hep-th/0206167}.


\bibitem{CK}
C.~S.~Chu and V.~V.~Khoze,
{\it ``Correspondence between the 3-point BMN correlators and the
  3-string  
vertex on the pp-wave,''} JHEP {\bf 0304} (2003) 014, 
{\tt hep-th/0301036}.

\bibitem{GK}
G.~Georgiou and V.~V.~Khoze,
{\it ``BMN operators with three scalar impurities and the
  vertex-correlator  duality in pp-wave,''} 
JHEP {\bf 0304} (2003) 015, 
{\tt hep-th/0302064}.

\bibitem{gursoy}
U.~Gursoy,
{\it ``Vector operators in the BMN correspondence,''}
{\tt hep-th/0208041}. 


\bibitem{beisert}
N.~Beisert,
{\it ``BMN operators and superconformal symmetry,''}
{\tt hep-th/0211032}.
  

\bibitem{klose}
T.~Klose,
{\it ``Conformal dimensions of two-derivative BMN operators,''}
JHEP {\bf 0303} (2003) 012,
{\tt hep-th/0301150}.


\bibitem{z2-1}
C.~S.~Chu, V.~V.~Khoze, M.~Petrini, R.~Russo and A.~Tanzini,
{\it ``A note on string interaction on the pp-wave background,''}
{\tt hep-th/0208148}.

\bibitem{z2-2}
C.~S.~Chu, M.~Petrini, R.~Russo and A.~Tanzini,
{\it ``String interactions and discrete symmetries of the pp-wave background,''}
{\tt hep-th/0211188}.


\bibitem{gross}  
D.~J.~Gross, A.~Mikhailov and R.~Roiban,  
{\it ``Operators with large R charge in N = 4 Yang-Mills theory,'' } 
Annals Phys.\  {\bf 301} (2002) 31, {\tt hep-th/0205066}.  

\bibitem{zanon}  
A.~Santambrogio and D.~Zanon,  
{\it ``Exact anomalous dimensions of {\cal N}=4 Yang-Mills operators with  
large R charge,''} Phys.\ Lett.\ B {\bf 545} (2002) 425,
{\tt hep-th/0206079}.  




\bibitem{KPSS}  
C.~Kristjansen, J.~Plefka, G.~W.~Semenoff and M.~Staudacher,  
{\it ``A new double-scaling limit of N = 4 super Yang-Mills theory and  
PP-wave  strings,''}  Nucl.\ Phys.\ B {\bf 643} (2002) 3,  
{\tt hep-th/0205033}.  
  




\bibitem{gross2}
D.~J.~Gross, A.~Mikhailov and R.~Roiban, 
{\it ``A calculation of the
plane wave string Hamiltonian from N = 4  super-Yang-Mills
theory,''} 
{\tt hep-th/0208231}.

\bibitem{ver}
H.~Verlinde,  
{\it ``Bits, matrices and 1/N,''}  
{\tt hep-th/0206059}.  
  


\bibitem{bits2}
D.~Vaman and H.~Verlinde,
{\it ``Bit strings from N = 4 gauge theory,''}
{\tt hep-th/0209215};\\
J.~Pearson, M.~Spradlin, D.~Vaman, H.~Verlinde and A.~Volovich,
{\it ``Tracing the string: BMN correspondence at finite $J^2/N$,''}
{\tt hep-th/0210102}.

\bibitem{zhou}
J.~G.~Zhou,
{\it ``pp-wave string interactions from string bit model,''}
Phys.\ Rev.\ D {\bf 67} (2003) 026010,
{\tt hep-th/0208232.}


\bibitem{Gomis}
J.~Gomis, S.~Moriyama and J.~w.~Park,
{\it ``SYM description of SFT Hamiltonian in a pp-wave 
background,''}
{\tt hep-th/0210153;} 
{\it ``SYM description of pp-wave string interactions: Singlet sector and  
arbitrary impurities,''} 
{\tt hep-th/0301250.}






\bibitem{Bianchi}
M.~Bianchi, B.~Eden, G.~Rossi and Y.~S.~Stanev,
{\it ``On operator mixing in N = 4 SYM,''}
Nucl.\ Phys.\ B {\bf 646} (2002) 69,
{\tt hep-th/0205321}.



\bibitem{Fubini}
S.~Fubini, A.~J.~Hanson and R.~Jackiw,
{\it ``New Approach To Field Theory,''}
Phys.\ Rev.\ D {\bf 7} (1973) 1732.

\bibitem{mack}
G.~Mack and A.~Salam,
{\it ``Finite Component Field Representations Of The Conformal Group,''}
Annals Phys.\  {\bf 53} (1969) 174.

\bibitem{osborn}
H.~Osborn and A.~C.~Petkou,
{\it ``Implications Of Conformal Invariance In Field Theories 
For General Dimensions,''}
Annals Phys.\  {\bf 231} (1994) 311,
{\tt hep-th/9307010}.

\bibitem{Fradkin}
E.~S.~Fradkin and M.~Y.~Palchik,
{\it ``Conformal Quantum Field Theory In D-Dimensions,''}
Dordrecht, Netherlands: Kluwer (1996)  
(Mathematics and its applications, volume 376).




\bibitem{D'Hoker:1998tz}
E.~D'Hoker, D.~Z.~Freedman and W.~Skiba,
{\it ``Field theory tests for correlators in the AdS/CFT correspondence,''}
Phys.\ Rev.\ D {\bf 59} (1999) 045008,
{\tt hep-th/9807098}.



  
  
  
  
\end{thebibliography}
\end{document}